%% file: BinaryDihedral.tex
\documentclass[a4paper,11pt]{article}
\usepackage[DIV12]{typearea}
\usepackage{longtable}
\usepackage{hhline}
\usepackage{booktabs,colortbl}
\usepackage{multirow}
\usepackage{amsmath}
\usepackage{amssymb}
\usepackage{bbm}
\usepackage{bm}
\usepackage{comment}
\usepackage[utf8]{inputenc}
\usepackage{pdflscape}
\usepackage{xspace}
\usepackage[caption=false]{subfig}
\usepackage{nicefrac}
\usepackage{graphicx}
\usepackage{adjustbox}
\usepackage{cite}  
\usepackage[small]{caption2}
\usepackage{mathtools}
\usepackage{nccfoots}
\usepackage{tikz}
\usepackage{enumitem}
\usepackage{blkarray, bigstrut}
\usepackage{braket}
\allowdisplaybreaks[3]

\usetikzlibrary{calc,positioning}
\usepackage{pgfplots}%
\pgfplotsset{compat = newest}

\usepackage[all,cmtip]{xy}
\newlength{\xywd}
\newcommand{\xyrightarrow}[2][]{%
\sbox{0}{$\scriptstyle#1$}%
\xywd=\wd0
\sbox{0}{$\scriptstyle#2$}%
\ifdim\wd0>\xywd \xywd=\wd0 \fi
\xymatrix@C\dimexpr\xywd+1em\relax{{}\ar[r]^{#2}_{#1}&{}}%
}

\DeclareMathOperator{\re}{Re}
\DeclareMathOperator{\im}{Im}

\newcommand{\rep}[1]{\ensuremath\boldsymbol{#1}}

\newcommand{\Z}[1]{\ensuremath{\mathbbm{Z}_{#1}}} % z_N ->\Z{N}
\newcommand{\SO}[1]{\ensuremath{\mathrm{SO}(#1)}}
\newcommand{\SU}[1]{\ensuremath{\mathrm{SU}(#1)}}

\newcommand{\e}{\mathrm{e}}
\newcommand{\dd}{\mathrm{d}}
\newcommand{\I}{\mathrm{i}}
\newcommand{\Id}{\mathbbm{1}}

\newcommand{\CP}{\ensuremath{\mathcal{CP}}\xspace}
\newcommand{\x}{\ensuremath{\times}}
\newcommand{\vev}[1]{\ensuremath{\langle{#1}\rangle}}

\usepackage{xcolor}
\definecolor{darkgreen}{HTML}{109930}
\definecolor{pink}{rgb}{0.858, 0.188, 0.478}

\newcommand{\ignore}[1]{}

\newcommand{\MeV}{\mathrm{MeV}}
\newcommand{\meV}{\mathrm{meV}}

\newcommand{\pa}{\bm{1}}
\newcommand{\pb}{\bm{1'}}
\newcommand{\pc}{\bm{\hat{1}}}
\newcommand{\pd}{\bm{\hat{1}'}}
\newcommand{\qa}{\bm{2}}
\newcommand{\qb}{\bm{\hat{2}}}
\newcommand{\MDDi}{\begin{array}{cc}
-\alpha_1 Y^{(k_{DD})}_{\qa,1}+\alpha_2 Y^{(k_{DD})}_{\pa}
&
\alpha_1 Y^{(k_{DD})}_{\qa,2}+\alpha_3 Y^{(k_{DD})}_{\pb}
\\
\alpha_1 Y^{(k_{DD})}_{\qa,2}-\alpha_3 Y^{(k_{DD})}_{\pb}
&
\alpha_1 Y^{(k_{DD})}_{\qa,1}+\alpha_2 Y^{(k_{DD})}_{\pa}
\\
\end{array}}

\newcommand{\MDDii}{\begin{array}{cc}
\alpha_1 Y^{(k_{DD})}_{\qb,2}+\alpha_3 Y^{(k_{DD})}_{\pd}
&
\alpha_1 Y^{(k_{DD})}_{\qb,1}+\alpha_2 Y^{(k_{DD})}_{\pc}
\\
\alpha_1 Y^{(k_{DD})}_{\qb,1}-\alpha_2 Y^{(k_{DD})}_{\pc}
&
-\alpha_1 Y^{(k_{DD})}_{\qb,2}+\alpha_3 Y^{(k_{DD})}_{\pd}
\\
\end{array}}

\newcommand{\MDDiib}{\begin{array}{cc}
\pm\alpha_1 Y^{(k_{DD})}_{\qb,2}+\alpha_3 Y^{(k_{DD})}_{\pd}
&
\pm\alpha_1 Y^{(k_{DD})}_{\qb,1}+\alpha_2 Y^{(k_{DD})}_{\pc}
\\
\pm\alpha_1 Y^{(k_{DD})}_{\qb,1}-\alpha_2 Y^{(k_{DD})}_{\pc}
&
\mp\alpha_1 Y^{(k_{DD})}_{\qb,2}+\alpha_3 Y^{(k_{DD})}_{\pd}
\\
\end{array}}

\newcommand{\MDDiii}{\begin{array}{cc}
\alpha_1 Y^{(k_{DD})}_{\qa,2}+\alpha_3 Y^{(k_{DD})}_{\pb}
&
\alpha_1 Y^{(k_{DD})}_{\qa,1}+\alpha_2 Y^{(k_{DD})}_{\pa}
\\
\alpha_1 Y^{(k_{DD})}_{\qa,1}-\alpha_2 Y^{(k_{DD})}_{\pa}
&
-\alpha_1 Y^{(k_{DD})}_{\qa,2}+\alpha_3 Y^{(k_{DD})}_{\pb}
\\
\end{array}}

\newcommand{\MDDiiib}{\begin{array}{cc}
\pm\alpha_1 Y^{(k_{DD})}_{\qa,2}+\alpha_3 Y^{(k_{DD})}_{\pb}
&
\pm\alpha_1 Y^{(k_{DD})}_{\qa,1}+\alpha_2 Y^{(k_{DD})}_{\pa}
\\
\pm\alpha_1 Y^{(k_{DD})}_{\qa,1}-\alpha_2 Y^{(k_{DD})}_{\pa}
&
\mp\alpha_1 Y^{(k_{DD})}_{\qa,2}+\alpha_3 Y^{(k_{DD})}_{\pb}
\\
\end{array}}

\newcommand{\MDDiv}{\begin{array}{cc}
-\alpha_1 Y^{(k_{DD})}_{\qb,1}+\alpha_3 Y^{(k_{DD})}_{\pc}
&
\alpha_1 Y^{(k_{DD})}_{\qb,2}+\alpha_2 Y^{(k_{DD})}_{\pd}
\\
\alpha_1 Y^{(k_{DD})}_{\qb,2}-\alpha_2 Y^{(k_{DD})}_{\pd}
&
\alpha_1 Y^{(k_{DD})}_{\qb,1}+\alpha_3 Y^{(k_{DD})}_{\pc}
\\
\end{array}}

\newcommand{\MDTi}{\hspace{-10pt}\begin{array}{c}
\beta Y^{(k_{D3})}_{\qa,1}
\\
\beta Y^{(k_{D3})}_{\qa,2}
\\
\end{array}}

\newcommand{\MDTii}{\hspace{-10pt}\begin{array}{c}
-\beta Y^{(k_{D3})}_{\qa,2}
\\
\beta Y^{(k_{D3})}_{\qa,1}
\\
\end{array}}

\newcommand{\MDTiib}{\hspace{-10pt}\begin{array}{c}
\mp\beta Y^{(k_{D3})}_{\qa,2}
\\
\pm\beta Y^{(k_{D3})}_{\qa,1}
\\
\end{array}}

\newcommand{\MDTiii}{\hspace{-10pt}\begin{array}{c}
\beta Y^{(k_{D3})}_{\qb,2}
\\
-\beta Y^{(k_{D3})}_{\qb,1}
\\
\end{array}}

\newcommand{\MDTiiib}{\hspace{-10pt}\begin{array}{c}
\pm\beta Y^{(k_{D3})}_{\qb,2}
\\
\mp\beta Y^{(k_{D3})}_{\qb,1}
\\
\end{array}}

\newcommand{\MDTiv}{\hspace{-10pt}\begin{array}{c}
\beta Y^{(k_{D3})}_{\qb,1}
\\
\beta Y^{(k_{D3})}_{\qb,2}
\\
\end{array}}

\newcommand{\MTDi}{\begin{array}{cc}
\gamma Y^{(k_{3D})}_{\qa,1} ~~~~~~~~~~~ &
~~~~~~~~~~~ \gamma Y^{(k_{3D})}_{\qa,2}
\\
\end{array}}

\newcommand{\MTDii}{\begin{array}{cc}
-\gamma Y^{(k_{3D})}_{\qa,2} ~~~~~~~~~~~ &
~~~~~~~~~~~ \gamma Y^{(k_{3D})}_{\qa,1}
\\
\end{array}}

\newcommand{\MTDiii}{\begin{array}{cc}
\gamma Y^{(k_{3D})}_{\qb,2} ~~~~~~~~~~~ &
~~~~~~~~~~~ -\gamma Y^{(k_{3D})}_{\qb,1}
\\
\end{array}}

\newcommand{\MTDiv}{\begin{array}{cc}
\gamma Y^{(k_{3D})}_{\qb,1} ~~~~~~~~~~~ &
~~~~~~~~~~~ \gamma Y^{(k_{3D})}_{\qb,2}
\\
\end{array}}

\newcommand{\MTDibis}{\begin{array}{cc}
		\gamma Y^{(k_{3D})}_{\qa,1} &
		\gamma Y^{(k_{3D})}_{\qa,2}
		\\
\end{array}}

\newcommand{\MTDiibbis}{\begin{array}{cc}
		\mp\gamma Y^{(k_{3D})}_{\qa,2} &
		\pm\gamma Y^{(k_{3D})}_{\qa,1}
		\\
\end{array}}

\newcommand{\MTDiiibbis}{\begin{array}{cc}
		\pm\gamma Y^{(k_{3D})}_{\qb,2} &
		\mp\gamma Y^{(k_{3D})}_{\qb,1}
		\\
\end{array}}

\newcommand{\MTDivbis}{\begin{array}{cc}
		\gamma Y^{(k_{3D})}_{\qb,1} &
		\gamma Y^{(k_{3D})}_{\qb,2}
		\\
\end{array}}

\newcommand{\MTTi}{\hspace{-10pt}\delta Y^{(k_{33})}_{\pa}}

\newcommand{\MTTii}{\hspace{-10pt}\delta Y^{(k_{33})}_{\pb}}

\newcommand{\MTTiii}{\hspace{-10pt}\delta Y^{(k_{33})}_{\pc}}

\newcommand{\MTTiv}{\hspace{-10pt}\delta Y^{(k_{33})}_{\pd}}

\advance \headheight by 3.0truept       % for 12pt mandatory...
\usepackage[pdftex]{hyperref}
\hypersetup{
pdftitle = {Quark and lepton modular models from the binary dihedral flavor symmetry},
pdfauthor = {Arriaga-Osante, Liu, Ramos-Sanchez}
}

\begin{document}

\begin{titlepage}

%\vspace*{-3.0cm}
\begin{flushright}
\end{flushright}

\vspace*{1.0cm}

\begin{center}
{\Large\textbf{Quark and lepton modular models\\ from the binary dihedral flavor symmetry}}

\vspace{1cm}
\textbf{Carlos Arriaga--Osante}$^1$,
\textbf{Xiang-Gan Liu}$^2$, and
\textbf{Sa\'ul Ramos--S\'anchez}$^1$
\Footnote{*}{%
	\href{mailto:cebrinarriaga@ciencias.unam.mx;xianggal@uci.edu;ramos@fisica.unam.mx}{\tt email addresses}
}
\\[5mm]
$^1$\textit{\small Instituto de F\'isica, Universidad Nacional Aut\'onoma de M\'exico,\\ Cd.~de M\'exico C.P.~04510, M\'exico}\\
$^2$\textit{\small Department of Physics and Astronomy, University of California, Irvine, CA 92697-4575, USA}
\end{center}

\vspace{1cm}

\vspace*{1.0cm}

\begin{abstract}
Inspired by the structure of top-down derived models endowed with modular flavor symmetries, we investigate the yet phenomenologically unexplored binary dihedral group $2D_3$. After building the vector-valued modular forms in the representations of $2D_3$ with small modular weights, we systematically classify all (Dirac and Majorana) mass textures of fermions with fractional modular weights and all possible $2+1$-family structures. This allows us to explore the parameter space of fermion models based on $2D_3$, aiming at a description of both quarks and leptons with a minimal number of parameters and best compatibility with observed data. We consider the separate possibilities of neutrino masses generated by either a type-I seesaw mechanism or the Weinberg operator.
We identify a model that, besides fitting all known flavor observables, delivers predictions for six not-yet measured parameters and favors normal-ordered neutrino masses generated by the Weinberg operator. It would be interesting to figure out whether it is possible to embed our model within a top-down scheme, such as $\mathbbm{T}^2/\Z4$ heterotic orbifold compactifications.
\end{abstract}

\end{titlepage}

\clearpage

{\hypersetup{linkcolor=black}
\tableofcontents
}

%%%%%%%%%%%%%%%%%%%%%%%%%%%%%%%%%%%%%%%%%%%%%%%%%%%%%%%%%%
\section{Introduction}
\label{sec:Introduction}

\input{Introduction.tex}

%%%%%%%%%%%%%%%%%%%%%%%%%%%%%%%%%%%%%%%%%%%%%%%%%%%%%%%%%%
\section{Modular symmetry and vector-valued modular forms }
\label{sec:VVMF}

\input{VVMF.tex}

%%%%%%%%%%%%%%%%%%%%%%%%%%%%%%%%%%%%
\section{Binary dihedral group of order 12}
\label{sec:2D3group}

\input{2D3.tex}

%%%%%%%%%%%%%%%%%%%%%%%%%%%%%%%%%%%%
\section{VVMFs of the modular binary dihedral group}
\label{sec:VVMF-modules-2D3}

\input{VVMFin2D3.tex}

%%%%%%%%%%%%%%%%%%%%%%%%%%%%%%%%%%%%%
\section{\boldmath Mass textures in models with $2D_3$ modular flavor\unboldmath}
\label{sec:model}

\input{classification.tex}

%%%%%%%%%%%%%%%%%%%%%%%%%%%%%%%%%%%%
\section{Benchmark models and numerical analysis}
\label{sec:BenchmarkModels}

\input{models.tex}

%%%%%%%%%%%%%%%%%%%%%%%%%%%%%%%%%%%%
\section{Conclusions}
\label{sec:conclusion}

\input{Conclusion.tex}

%%%%%%%%%%%%%%%%%%%%%%%%%%%%%%%%%%%%%%%%%%%%%%%%%%%%%%%%%%%%%%%%%%%%%%%%%%%%%%%%%%%%%%%%%%%%%%%%%%%%%%%%%%%%%%
\section*{Acknowledgments}
It is a pleasure to thank Ram\'on D\'iaz-Castro for useful discussions during this work.
We also thank Alexander Baur, Hans Peter Nilles, Patrick K.S.~Vaudrevange and Andreas Trautner for 
enlightening top-down discussions. This work is partially supported by UC-MEXUS-CONACyT grant No.\ CN-20-38.
CAO and SRS are supported by UNAM-PAPIIT IN113223, CONACyT grant CB-2017-2018/A1-S-13051, and Marcos Moshinsky Foundation. 
XGL is supported by U.S.\ National Science Foundation under Grant No.~PHY-2210283.

%%%%%%%%%%%%%%%%%%%%%%%%%%%%%%%%%%%%%%%%%%%%%%%%%%%%%%%%%%%%%%%%%%%%%%%%%%%%%%%%%%%%%%%%%%%%%%%%%%%%%%%%%%%%%%

\newpage
\begin{appendix}

%%%%%%%%%%%%%%%%%%%%%%%%%%%%%%%%%%%%
\section{Generalized hypergeometric series}
\label{app:hypergeometric}

\input{app_hypergeometric.tex}

%%%%%%%%%%%%%%%%%%%%%%%%%%%%%%%%%%%%
\section{Clasification of Dirac mass matrices}
\label{app:classification_Dirac}

\input{app_Dirac.tex}

\newpage
%%%%%%%%%%%%%%%%%%%%%%%%%%%%%%%%%%%%
\section{Classification of Majorana mass matrices}
\label{app:classification_Majorana}

\input{app_Majorana.tex}

\end{appendix}

%%%%%%%%%%%%%%%%%%%%%%%%%%%%%%%%%%%%%%%%%%%%%%%%%%%%%%%%%%%%%%%%%%%%%%%%%%%%%%%%%%%%%%%%%%%%%%%%%%%%%%%%%%%%%%
{\small
%\bibliographystyle{OurBibTeX}
%\bibliography{references}

\providecommand{\bysame}{\leavevmode\hbox to3em{\hrulefill}\thinspace}

}
\end{document}

%% file: Introduction.tex
One of the extant questions of the Standard Model (SM) is the flavor puzzle: what is the origin of the observed multiplicity of matter generations or flavors, and their masses and mixings? It is conceivable that the answer is associated with the existence of a symmetry among flavors, likely encoded in a non-Abelian finite group (see e.g.~\cite{Altarelli:2010gt,Ishimori:2010au,King:2013eh,Feruglio:2019ybq,Chauhan:2023faf}). In his seminal work~\cite{Feruglio:2017spp}, Feruglio showed that it is possible to build phenomenologically viable extensions of the SM in which such a symmetry is modular, implying that couplings are modular forms that depend only on a complex modulus field $\tau$. Interestingly, flavons are not mandatory in this formalism as the vacuum expectation value (VEV) of $\tau$ sets all observable properties. Hence, the flavor model's parameter space is greatly reduced with respect to scenarios based on non-modular or traditional flavor groups. Despite some challenges~\cite{Chen:2019ewa}, this idea triggered a vigorous bottom-up quest to extract predictions from promising models based mostly on finite modular groups of the type $\Gamma_N := \mathrm{PSL}(2,\Z{}) / \overline\Gamma(N)$ or $\Gamma'_N := \mathrm{SL}(2,\Z{}) / {\Gamma} (N)$, where $\overline\Gamma(N)$ ($\Gamma(N)$) is a congruence subgroup of level $N$ of $\mathrm{PSL}(2,\Z{})$ ($\mathrm{SL}(2,\Z{})$). In this framework, several models fitting observable data have been identified. E.g.\ for $N\leq7$, we find models based on $\Gamma_2 \cong S_3$ \cite{Kobayashi:2018vbk,Kobayashi:2018wkl}, $\Gamma_3 \cong A_4$ \cite{Ding:2019zxk,Feruglio:2017spp,Kobayashi:2018scp,Kobayashi:2018wkl,Novichkov:2018yse}, $\Gamma_4 \cong S_4$ \cite{Criado:2019tzk,Kobayashi:2019mna,Novichkov:2018ovf,Penedo:2018nmg}, $\Gamma_5 \cong A_5$ \cite{Criado:2019tzk,Ding:2019xna,Novichkov:2018nkm}, $\Gamma_6\cong S_3 \times A_4$ \cite{Kikuchi:2023cap},  $\Gamma_7 \cong \text{PSL}(2,\Z7)$ \cite{Ding:2020msi}, $\Gamma'_3 \cong T'$ \cite{Liu:2019khw,Lu:2019vgm}, $\Gamma'_4 \cong S'_4$ \cite{Liu:2020akv,Novichkov:2020eep}, $\Gamma'_5 \cong A'_5$ \cite{Wang:2020lxk}, $\Gamma'_6 \cong S_3 \times T'$ \cite{Li:2021buv}. These models rely on a clever choice of representations and modular weights for both the modular forms associated with the couplings and the matter fields, as well as on the VEV of $\tau$, whose stabilization mechanism deserves further analysis~\cite{Ishiguro:2020tmo,Novichkov:2022wvg,Ishiguro:2022pde,Knapp-Perez:2023nty,King:2023snq,Kobayashi:2023spx,McAllister:2023vgy}. Besides successful fits to flavor observables, including texture zeroes~\cite{Zhang:2019ngf,Lu:2019vgm,Kikuchi:2022svo,Ding:2022aoe,Nomura:2023usj}, this scheme suggests that the origin of flavor hierarchies might be the breakdown of modular symmetries close to fixed points in moduli space~\cite{Ding:2019gof,Feruglio:2021dte,Kikuchi:2023cap,King:2019vhv,Kobayashi:2021pav,Novichkov:2018yse,Novichkov:2021evw,Okada:2020ukr}.

Beyond $\Gamma_N$ and $\Gamma'_N$ modular symmetries, the bottom-up modular framework can be straightforwardly extended to metaplectic~\cite{Liu:2020msy,Yao:2020zml} and symplectic groups~\cite{Ding:2020zxw}, as well as the resulting quotients from dividing $\mathrm{PSL}(2,\Z{})$ or $\mathrm{SL}(2,\Z{})$ by any of their various normal subgroups~\cite{Liu:2021gwa,Ding:2023ydy}, all supporting generalized vector-valued modular forms (VVMFs). This enlarges the possibilities for a modular theory of flavor. However, it could also be considered a challenge in bottom-up constructions as there has not been identified a way to single out the right set of modular symmetry, representations and modular weights for the elements of the models.

On the other hand, additional motivation for modular symmetries comes from their natural appearance in top-down constructions~\cite{Lauer:1989ax,Lauer:1990tm,Ibanez:1992hc} and their direct interpretation as flavor symmetries~\cite{Kobayashi:2018bff,Kobayashi:2018rad,Baur:2019kwi,Kikuchi:2020frp,Kikuchi:2020nxn,Ohki:2020bpo,Almumin:2021fbk}. This typically includes a generalized \CP-like modular transformation that accompanies the flavor symmetry~\cite{Baur:2019iai} (which has been studied from a bottom-up viewpoint too~\cite{Novichkov:2019sqv,Ding:2021iqp}). Contrasting with bottom-up constructions where global supersymmetry (SUSY) is mostly chosen, top-down flavor models offer a full scheme within supergravity (SUGRA), in which the K\"ahler potential transforms non-trivially and the superpotential carries a modular weight to yield a modular invariant theory. Among the top-down constructions, $\mathbbm{T}^2/\Z{N}$ two-dimensional (2-d) toroidal heterotic orbifold compactifications have shown to lead to simple yet promising modular scenarios~\cite{Nilles:2020kgo,Nilles:2020tdp,Nilles:2020gvu,Baur:2022hma}. In the cases studied so far, the flavor symmetry comprises a modular component plus a traditional geometry-based flavor subgroup~\cite{Kobayashi:2006wq}, building a so-called eclectic flavor group~\cite{Nilles:2020nnc}. 

A great advantage of this top-down formalism is that string properties fully define not only the flavor group, but also other aspects, such as the modular representations and weights for matter fields and the modular features of the couplings among them. The simplest top-down scenario emerges from a $\mathbbm{T}^2/\Z3$ orbifold, which leads to an eclectic group~\cite{Nilles:2020tdp} built from a $\Gamma_3'\cong T'$ modular symmetry combined with a traditional $\Delta(54)$ flavor group~\cite{Baur:2019kwi,Baur:2019iai}. The three SM generations in this case are accommodated in $\rep2'\oplus\rep1$ or $\rep2''\oplus\rep1$ representations of $T'$ instead of the apparently more natural triplets~\cite{Nilles:2020kgo}, and the Higgs is a trivial or non-trivial singlet~\cite{Baur:2022hma}. The $2+1$-family structure seems to be quite generic in these top-down constructions~\cite{Nilles:2004ej,Lebedev:2007hv,Nilles:2008gq}, and is phenomenologically favored given the known mixings and mass textures of observed fermions. Furthermore, top-down models of flavor exhibit multiples of $\nicefrac13$ as modular weights for matter generations~\cite{Ibanez:1992hc,Olguin-Trejo:2017zav,Nilles:2020tdp} while the Higgs fields enjoy vanishing or interger weights.

Additionally, $\mathbbm{T}^2/\Z2$ heterotic orbifolds yield an eclectic structure that include a $\Gamma'_2\cong S_3$ modular flavor component, fermions with multiples of $\nicefrac12$ as modular weights~\cite{Baur:2020jwc,Baur:2021mtl}. Following this pattern, it is expected that modular flavor symmetries from $\mathbbm{T}^2/\Z{N}$ models are associated in general with $\Gamma'_N$ groups. After \Z2 and \Z3 orbifolds, the next simple case arises from $\mathbbm{T}^2/\Z4$ orbifolds, which may contain a $\Gamma'_4\cong S_4'\cong[48,30]$ modular flavor group. (We use the Small-Groups library GAP Id~\cite{GAP}, where the first number in the square brakets denotes the order of the group.) A careful top-down study of this case reveals that only its quotient group $S_4'/\Z2\x\Z2 \cong 2D_3\cong[12,1]$ is the actual modular flavor group respected by effective matter fields at the massless level~\cite{Baur:2023}. Interestingly, the binary dihedral group $2D_3$ corresponds to the second smallest possible finite modular group~\cite{Liu:2021gwa} and its flavor phenomenology remains unexplored. Further, the two doublets and four singlets of $2D_3$ suggest the $2+1$ promising structure for matter generations.

Inspired by these observations, one of the goals of this work is to arrive at possible bottom-up realizations of viable models based on the modular flavor symmetry $2D_3$, exhibiting some of the top-down features: i) $2+1$-family structure, ii) fractional modular weights for fermions as multiples of $\nicefrac14$,
and iii) Higgs fields building singlet representations with vanishing weights and singlet representations. Our models should be {\it complete} in the sense that they must not only comply with observable constraints on leptons alone, but also on the quark sector, which has revealed to be more challenging to be fit in the modular framework, see e.g.~\cite{deAnda:2018ecu,Okada:2018yrn,Kobayashi:2019rzp,Lu:2019vgm,King:2020qaj,Okada:2020ukr,Ding:2020zxw,Zhao:2021jxg,Chen:2021zty,Ding:2021eva,Kikuchi:2023jap,Kikuchi:2023dow}. Hence, we aim at providing the vector-valued modular forms corresponding to our chosen group, classify the possible mass and mixings, and identify a complete fit of both the quark and lepton sectors.

This paper is organized as follows. In section~\ref{sec:VVMF} we review the theory of vector-valued modular forms~\cite{Liu:2021gwa}. Section~\ref{sec:2D3group} contains the group theory of the binary dihedral group whereas in section~\ref{sec:VVMF-modules-2D3} we identify the vector-valued modular forms under the irreducible representations of the binary dihedral group up to weight 7. Section~\ref{sec:model} is devoted to the invariant theory under the group $2D_3$ and a systematic classification of mass textures for the whole fermion sector considering all possible Dirac and Majorana mass matrices, associated with neutrino masses generated either by a type-I seesaw mechanism or the Weinberg operator. Finally, in section~\ref{sec:BenchmarkModels} we present the results of our systematic numerical analysis leading to the best-fit complete models for the quark and lepton sectors separately, and a complete model for quarks and leptons together.

%% file: VVMF.tex
Modular theories of flavor are mostly based on the modular group $\Gamma:=\mathrm{SL}(2,\Z{})$. While in bottom-up models the appearance of this group can be just assumed, in top-down constructions $\Gamma$ can arise from the geometric symmetries of a 2-d torus $\mathbbm{T}^2$ on which two extra dimensions are compactified. Regardless of its origin, $\Gamma$ can be defined by
\begin{equation}
 \Gamma~=~\left\{\begin{pmatrix}
    a &~ b \\ c &~ d
 \end{pmatrix}  ~\Big|~ ad- bc=1\,,\quad  a,b,c,d \in \Z{}\,\right\}\,.
\end{equation}
This group can be generated by two elements $S$ and $T$ obeying the relations
\begin{equation}
\label{eq:SL2Zrelation}
  S^4~=~(ST)^3~=~\Id\,,\qquad 
  S^2 T~=~T S^2\,,
\end{equation}
which can be represented as
\begin{equation}
  S~=~\begin{pmatrix}
    0 &~ 1 \\ -1 &~ 0
  \end{pmatrix}
  \qquad \text{and} \qquad
  T~=~\begin{pmatrix}
    1 &~ 1 \\ 0 &~ 1
  \end{pmatrix}\,.
\end{equation}
Note that $S^2=-\Id_2$, where $\Id_2$ denotes the 2-d identity matrix. 
The moduli space of a $\mathbbm{T}^2$ is spanned by a modulus $\tau$, which lies in the complex upper-half plane 
$\mathcal{H}=\{\tau\in\mathbb{C}~|~\text{Im}\tau >0\}$. 
The modular group acts on the moduli space $\mathcal{H}$ through the linear fractional transformation
\begin{equation}
\label{eq:tautransformation}
  \tau ~\rightarrow~ \gamma \tau ~:=~\frac{a\tau+b}{c\tau+d}\qquad\text{with}\qquad 
  \gamma~=~\begin{pmatrix}
           a & b \\c & d
           \end{pmatrix}\in \Gamma\,.
\end{equation}
We note that, despite $\gamma$ belonging to $\Gamma$, its action on $\tau$ is equivalent to the action of $-\gamma$. This implies that that the modulus only perceives the action of the projective modular group $\overline\Gamma := \Gamma/\pm\Id$.

A special class of vector-valued holomorphic functions $Y(\tau)=(Y_1(\tau),\dots,Y_d(\tau))^{\rm T}$  exist in the extended upper-half plane $\mathcal{H}$ (including the point $\tau\to\I\infty$), exhibiting the non-trivial modular transformation under $\gamma\in\Gamma$
\begin{equation}
  \label{eq:defVVMF}
  Y(\tau)~\stackrel{\gamma}{\longrightarrow}~ Y(\gamma\tau)~=~(c\tau+d)^{k_{Y}}\rho_{Y}(\gamma)\,Y(\tau)\,,
\end{equation}
where $k_{Y}$ is a positive integer referred to as the modular weight of $Y(\tau)$, and $\rho_{Y}$ denotes a $d$-dimensional representation of $\gamma\in\Gamma$ (with finite image). Such holomorphic functions are known as vector-valued modular forms (VVMFs) or modular form multiplets, and they play a central role in the theory of modular flavor symmetries.

All of the VVMFs in 1-d irreducible representations (irreps) encompass the Eisenstein series $E_4(\tau)$, $E_6(\tau)$ and eta products~\cite{cohen2017modular}:
\begin{subequations}
\label{eq:E4-E6-eta} 
\begin{align}
  E_4(\tau)&=~1+240\sum_{n=1}^{\infty}\sigma_3(n)q^n \,,\\
  E_6(\tau)&=~1-504\sum_{n=1}^{\infty}\sigma_5(n)q^n\,, \\
  \eta^{2p}(\tau)&=~q^{p/12}\prod_{n=1}^\infty \left(1-q^n \right)^{2p}\,,
  \quad\text{with}\quad q:= \e^{2 \pi \I\tau}\,,
\end{align}
\end{subequations}
where $\sigma_k(n) = \sum_{d|n} d^k$ is the sum of the $k$th power of the divisors of $n$,  $p\in \mathbbm{N}^+$ and $\eta(\tau)$ is the Dedekind-eta function.

It should be noted that all VVMFs in the irrep $\rho$ make up a free module (denoted by $\mathcal{M}(\rho)$) over the ring $\mathcal{M}(\rep1)=\mathbbm{C}[E_4,E_6]$, whose rank is exactly equal to the dimension of $\rho$.  The module $\mathcal{M}(\rho)$ always has one or more VVMF of minimal weight, which is uniquely determined by the representation matrix $\rho(T)$.  The basis of module $\mathcal{M}(\rho)$ can typically be obtained by applying the modular differential operators $D^n_k$ to these VVMFs of minimal weight. The modular differential operators $D^n_k$ are defined as
\begin{equation}
  D^{n}_k ~:=~ D_{k+2(n-1)}\circ D_{k+2(n-2)}\circ\dots \circ D_k\,,
\end{equation}
where the modular derivative is defined by
\begin{equation}
 D_k ~:=~ \frac{1}{2\pi\I} \frac{\dd}{\dd\tau}- \frac{kE_2(\tau)}{12}\,,\qquad k\in\mathbbm{N}^+\,,
\end{equation}
and $E_2(\tau)$ denotes the quasi-modular Eisenstein series~\cite{cohen2017modular},
\begin{equation}
\label{eq:E2}
 E_2(\tau)~=~1-24\sum_{n=1}^{\infty}\sigma_1(n)q^n \,.
\end{equation}
It is not difficult to show that the operator $D^n_k$ acting on a VVMF $Y(\tau)$ of weight $k$ gives a VVMF $D^n_kY(\tau)$ of higher weight $k+2n$ in the same representation~\cite{Bruinier2008The,franc2016hypergeometric,Liu:2021gwa}. 
For $d:=\dim\rho\leq 3$, its VVMF module contains only one VVMF of minimal modular weight, and the basis of the module comprises %is composed of 
the set $\{Y^{(k_0)},D_{k_0} Y^{(k_0)},\dots,D_{k_0}^{d-1} Y^{(k_0)}\}$, where $Y^{(k_0)}$ is the VVMF of minimal weight $k_0$.  The modular form multiplets of higher weight $k_0+2d$ can be written as a linear combinations of these bases on the ring $\mathcal{M}(\rep{1})=\mathbbm{C}[E_4,E_6]$, i.e.
\begin{equation}
\label{eq:MLDE}
 (D^d_{k_0}+M_4D_{k_0}^{d-2}+\dots+M_{2(d-1)}D_{k_0}+M_{2d})Y^{(k_0)}~=~0\,,
\end{equation}
where $M_k\in \mathbbm{C}[E_4,E_6]$ is the scalar modular form of weight $k$. 
Eq.~\eqref{eq:MLDE}  can at the same time be interpreted as a modular linear differential equation (MLDE) satisfied by the VVMF of minimal weight $k_0$, whose solution provides us with the specific form of $Y^{(k_0)}$. For all 1-d, 2-d, and 3-d irreps $\rho$, the solutions $Y^{(k_0)}$ can be expressed either as eta products or generalized hypergeometric series. A more comprehensive overview of the theory of VVMFs, including more specific results, 
can be found in Ref.~\cite{Liu:2021gwa}.

%% file: 2D3.tex
The binary dihedral group $2D_3$ (with GAP~\cite{GAP} Id $[12,1]$) also known as the dicyclic group $Dic_{3}$ of order 12, is the preimage of the dihedral group $D_3$ under the spin group double cover map $\SU2 \cong \mathrm{Spin}(3)\to \SO3$. Equivalently, $2D_3$ is also the lift of the dihedral group $D_3$ through the pin group double cover map $ \mathrm{Pin}\_(2)\to \mathrm{O}(2)$~\cite{nlab:dihedral_group}.
The group $2D_3$ can be generated by the modular generators $S$ and $T$ satisfying the relations~\cite{Liu:2021gwa}
\begin{equation}
\label{eq:2Ogroup}
 S^4~=~(ST)^3~=~S^2T^2~=~\Id\,,\qquad S^2T~=~TS^2\,.
\end{equation}
The cyclic group $\Z2^{S^2}=\left\{1, S^2\right\}=\left\{1, T^2\right\}$ is the center of $2D_3$, and the quotient group $2D_3/\Z2^{S^2}$ is isomorphic to $S_3$, which means that $2D_3$ can be regarded as a central extension of the finite modular group $\Gamma_2\cong S_3$.  Notice also that $2D_3\cong\Z3\rtimes\Z4$.

\begin{table}[t!]
{\centering
\begin{tabular}{|c||c|c|c|c|c|c|}
\hline
$\rep{r}$        & $\bm{1}$ & $\bm{1'}$ & $\bm{\hat{1}}$ & $\bm{\hat{1}'}$ & $\bm{2}$ & $\bm{\hat{2}}$ \\
\hline\hline
$\rho_{\rep{r}}(S)$ & $1$  & $-1$  &  $\I$   &  $-\I$ & $\dfrac{1}{2}\begin{pmatrix}-1 & \sqrt{3} \\ \sqrt{3}& 1 \end{pmatrix}$ 
& $\dfrac{\I}{2}\begin{pmatrix}-1 & \sqrt{3} \\ \sqrt{3}& 1 \end{pmatrix}$ \\
$\rho_{\rep{r}}(T)$ & $1$  & $-1$  &  $-\I$  &  $\I$  & $\begin{pmatrix}1 & 0 \\ 0 & -1 \end{pmatrix}$  
& $\begin{pmatrix}-\I & 0 \\ 0 & \I \end{pmatrix}$\\
\hline
\end{tabular}
\caption{Representations of the binary dihedral group $2D_3\cong \Z3\rtimes\Z4\cong[12,1]$.\label{tab:2D3reps}}
}
\end{table}

$2D_3$ possesses four singlets $\bm{1},\bm{1'},\bm{\hat{1}},\bm{\hat{1}'}$ and two doublet representations $\bm{2},\bm{\hat{2}}$. The explicit forms of the representation matrices $\rho_{\rep{r}}(S)$ and $\rho_{\rep{r}}(T)$ in these irreps are given in Table~\ref{tab:2D3reps}. As mentioned above, $2D_3/\Z2^{S^2}\cong S_3$. Hence, all the irreducible representations of $S_3$ are included in the list of representations of $2D_3$. They correspond to the representations $\rep{1}$, $\rep{1}'$ and $\rep{2}$ of Table~\ref{tab:2D3reps}. On the other hand, a similar relationship exists between $2D_3$ and the finite modular group $\Gamma'_4\cong S'_4$. In fact, $S'_4$ is the split extension of $2D_3$ by $\Z2\times \Z2$, i.e., $S'_4\cong (\Z2\times \Z2)\rtimes 2D_3$. Consequently, $2D_3$ is isomorphic to the quotient group $S'_4/(\Z2\x\Z2)$. This implies that all irreducible representations of $2D_3$ are also irreducible representations of $S'_4$. As one can verify, the above four singlets and two doublets coincide with the 1-d and 2-d irreps of $S'_4$~\cite{Novichkov:2020eep,Liu:2020akv,Ding:2022nzn}. Note that $2D_3$ is also a subgroup of $S'_4$.

The 12 elements of $2D_3$ can be divided into six conjugacy classes. The corresponding character table is shown in Table~\ref{tab:character}, which can be obtained, as usual, by taking the trace of the explicit representation matrices.

\begin{table}[t!]
\begin{center}
\renewcommand{\tabcolsep}{2.8mm}
\renewcommand{\arraystretch}{1.3}
\begin{tabular}{|c|c|c|c|c|c|c|c|c|c|c|c|c|c|c|}\hline\hline
\text{Classes} & $1C_1$ & $1C_2$ & $3C_4$ & $2C_3$ & $3C'_4$ & $2C_6$  \\ \hline
\text{Representative} & $1$ & $S^2$ & $T$ & $TS$ & $TS^2$ & $TS^3$ \\ \hline
$\bm{1}$ & $1$ & $1$ & $1$ & $1$ & $1$ & $1$\\ \hline
$\bm{1'}$ & $1$ & $1$ & $-1$ & $1$ & $-1$ & $1$\\ \hline
$\bm{\hat{1}}$ & $1$ & $-1$ & $-\I$ & $1$ & $\I$ & $-1$\\ \hline
$\bm{\hat{1}'}$ & $1$ & $-1$ & $\I$ & $1$ & $-\I$ & $-1$\\ \hline
$\bm{2}$ & $2$ & $2$ & $0$ & $-1$ & $0$ & $-1$ \\ \hline
$\bm{\hat{2}}$ & $2$ & $-2$ & $0$ & $-1$ & $0$ & $1$\\ \hline\hline
\end{tabular}
\caption{\label{tab:character} Character table of the binary dihedral group $2D_3\cong \Z3\rtimes\Z4\cong[12,1]$.}
\end{center}
\end{table}

The tensor products between singlets are given by
\begin{equation}
\bm{1'}\otimes\bm{1'}=\bm{1}\,,\quad \bm{1'}\otimes\bm{\hat{1}}=\bm{\hat{1}'}\,,\quad \bm{1'}\otimes\bm{\hat{1}'}=\bm{\hat{1}}\,,\quad \bm{\hat{1}}\otimes\bm{\hat{1}}=\bm{1'}\,,\quad 
\bm{\hat{1}}\otimes\bm{\hat{1}'}=\bm{1}\,,\quad 
\bm{\hat{1}'}\otimes\bm{\hat{1}'}=\bm{1'}\,.
\label{eq:SingletProductRules}
\end{equation}
The tensor products between singlet and doublet with components $(\beta_1,\beta_2)^\mathrm{T}$ are given by
\begin{subequations}
\begin{eqnarray}
\begin{cases}
\bm{1'}\otimes\bm{2}=\bm{2}\,,\quad \bm{\hat{1}'}\otimes\bm{2}=\bm{\hat{2}}\,\\ 
\bm{1'}\otimes\bm{\hat{2}}=\bm{\hat{2}}\,,\quad \bm{\hat{1}}\otimes\bm{\hat{2}}=\bm{2}\,
\end{cases} &:&
\qquad 
\alpha_1 \begin{pmatrix}
\beta_2 \\ - \beta_1
\end{pmatrix}\,,
\\
\begin{cases}
\bm{\hat{1}}\otimes\bm{2}=\bm{\hat{2}}\,,\quad \bm{\hat{1}'}\otimes\bm{\hat{2}}=\bm{2}\,\\
\end{cases} &:&
\qquad \alpha_1 \begin{pmatrix}
\beta_1 \\  \beta_2
\end{pmatrix}\,.
\end{eqnarray}
\end{subequations}
Finally, the tensor products between two doublets with respective components $(\alpha_1,\alpha_2)^\mathrm{T}$ and 
$(\beta_1,\beta_2)^\mathrm{T}$ are computed as
\begin{subequations}
\label{eq:DoubletProductRules}
\begin{eqnarray}
\begin{cases}
\bm{2}\otimes\bm{2}=\bm{1}\oplus\bm{1'}\oplus\bm{2}\\ \bm{2}\otimes\bm{\hat{2}}=\bm{\hat{1}}\oplus\bm{\hat{1}'}\oplus\bm{\hat{2}}\,
\end{cases} &:&
\left(\alpha_1\beta_1+\alpha_2\beta_2\right)  \oplus \left(\alpha_1\beta_2-\alpha_2\beta_1 \right) \oplus
\begin{pmatrix}
\alpha_2\beta_2-\alpha_1\beta_1 \\
\alpha_1\beta_2+\alpha_2\beta_1
\end{pmatrix}\,,
\\
\begin{cases}
\bm{\hat{2}}\otimes\bm{\hat{2}}=\bm{1}\oplus\bm{1'}\oplus\bm{2}\\
\end{cases} &:&
\left(\alpha_1\beta_2-\alpha_2\beta_1\right)  \oplus \left(\alpha_1\beta_1+\alpha_2\beta_2 \right) \oplus
\begin{pmatrix}
\alpha_1\beta_2+\alpha_2\beta_1 \\
\alpha_1\beta_1-\alpha_2\beta_2
\end{pmatrix}\,.
\end{eqnarray}
\end{subequations}

%% file: VVMFin2D3.tex
Since all the VVMFs (denoted by $\mathcal{M}(2D_3)$) in the representations of the binary dihedral group $2D_3$ are simply the set of each irreducible VVMF module, we can formally express $\mathcal{M}(2D_3)$ as the direct sum of six irreducible VVMFs modules,
\begin{equation}
 \mathcal{M}(2D_3)~=~\mathcal{M}(\bm{1})\oplus\mathcal{M}(\bm{1'})\oplus\mathcal{M}(\bm{\hat{1}})\oplus 
\mathcal{M}(\bm{\hat{1}'}) \oplus \mathcal{M}(\bm{2}) \oplus \mathcal{M}(\bm{\hat{2}})\,.
\end{equation}
Each $\mathcal{M}(\rho) = \bigoplus^{\infty}_{k=0}\mathcal{M}_k(\rho)$ in a representation $\rho$ is a graded module over the ring $\mathcal{M}(\bm{1})=\mathbbm{C}[E_4,E_6]$. A basis of the module $\mathcal{M}(\rho)$ is sufficient to generate the entire space of VVMFs. From the general theory of vector-valued modular forms~\cite{Liu:2021gwa}, we can determine the basis of each module and the VVMF of minimal weight within them. Specifically, each module can be generated by the following generators over the ring $\mathcal{M}(\bm{1})=\mathbbm{C}[E_4, E_6]$:
\begin{equation}
\label{eq:modules2D3} 
\begin{aligned}
  \mathcal{M}(\bm{1})&~=~\langle 1 \rangle\,,                            & \mathcal{M}(\bm{1'})&~=~\langle Y^{(6)}_{\bm{1'}} \rangle\,,\\
  \mathcal{M}(\bm{\hat{1}})&~=~\langle Y^{(9)}_{\bm{\hat{1}}} \rangle\,, & \mathcal{M}(\bm{\hat{1}'})&~=~\langle Y^{(3)}_{\bm{\hat{1}'}} \rangle\,,\\
  \mathcal{M}(\bm{2})&~=~\langle Y^{(2)}_{\bm{2}},~ D_2 Y^{(2)}_{\bm{2}} \rangle\,,\qquad &
  \mathcal{M}(\bm{\hat{2}})&~=~ \langle Y^{(5)}_{\bm{\hat{2}}},~ D_5Y^{(5)}_{\bm{\hat{2}}} \rangle\,.
\end{aligned}
\end{equation}
The methodology employed here to obtain the VVMFs of $2D_3$ with their admissible weights is similar to that described in Ref.~\cite{Liu:2021gwa} and will not be reiterated here. The resulting VVMFs of minimal weight read 
\begin{equation}
\label{eq:MinimalWeightVVMF2D3}
\begin{aligned}
Y^{(6)}_{\bm{1'}}(\tau)&~=~\eta^{12}(\tau)\,,\qquad Y^{(9)}_{\bm{\hat{1}}}(\tau)~=~\eta^{18}(\tau)\,,\qquad Y^{(3)}_{\bm{\hat{1}'}}(\tau)~=~\eta^{6}(\tau)\,, \\
Y^{(2)}_{\bm{2}}(\tau)&~=~\begin{pmatrix}
 \eta^{4}(\tau)\left(\frac{K(\tau)}{1728}\right)^{-\nicefrac16} {}_2F_1\left(-\frac16,\frac16;\frac12;K(\tau)\right) \\
 -8\sqrt{3}\eta^{4}(\tau)\left(\frac{K(\tau)}{1728}\right)^{\nicefrac13} {}_2F_1\left(\frac13,\frac23;\frac32;K(\tau)\right)
\end{pmatrix}\,, \\
Y^{(5)}_{\bm{\hat{2}}}(\tau)&~=~\begin{pmatrix}
 8\sqrt{3}\eta^{10}(\tau)\left(\frac{K(\tau)}{1728}\right)^{\nicefrac13} {}_2F_1\left(\frac13,\frac23;\frac32;K(\tau)\right) \\
 \eta^{10}(\tau)\left(\frac{K(\tau)}{1728}\right)^{-\nicefrac16} {}_2F_1\left(-\frac{1}{6},\frac{1}{6};\frac12;K(\tau)\right)
\end{pmatrix}\,,
\end{aligned}
\end{equation}
where ${}_2F_1$ is the generalized hypergeometric series and $K(\tau):=1728/j(\tau)$ is the inverse of Klein-$j$ function $j(\tau)$, as defined in Appendix~\ref{app:hypergeometric}.

It is interesting to note that the structure of the VVMFs in Eq.~\eqref{eq:MinimalWeightVVMF2D3} implies
\begin{equation}
Y^{(5)}_{\bm{\hat{2}}}(\tau) ~=~
 \left(Y^{(3)}_{\bm{1'}}(\tau)Y^{(2)}_{\bm{\hat{2}},2}(\tau),\quad -Y^{(3)}_{\bm{1'}}(\tau)Y^{(2)}_{\bm{\hat{2}},1}(\tau) \right)^\mathrm{T},
\end{equation}
which is exactly what we would expect from the tensor product $\bm{\hat{1}'}\otimes\bm{2}=\bm{\hat{2}}$.
The linearly independent modular form multiplets of $2D_3$ at each allowed weight can be obtained straightforwardly by multiplying the polynomial of $E_4,E_6$ with the bases of modules in Eq.~\eqref{eq:modules2D3}:
\begin{equation}
\label{eq:MF-multiplets-2D3}
\begin{aligned}
&k=2:~~ Y^{(2)}_{\bm{2}}\,,\\
&k=3:~~ Y^{(3)}_{\bm{\hat{1}'}}\,,\\
&k=4:~~ Y^{(4)}_{\bm{1}}:= E_4\,, \quad Y^{(4)}_{\bm{2}}:= -6D_2 Y^{(2)}_{\bm{2}}\,,\\
&k=5:~~ Y^{(5)}_{\bm{\hat{2}}}\,,\\
&k=6:~~ Y^{(6)}_{\bm{1}}:= E_6\,,\quad Y^{(6)}_{\bm{1}'}\,,\quad Y^{(6)}_{\bm{2}}:= E_4 Y^{(2)}_{\bm{2}}\,,\\
&k=7:~~ Y^{(7)}_{\bm{\hat{2}}}:= 6 D_5 Y^{(5)}_{\bm{\hat{2}}}\,,\quad Y^{(7)}_{\bm{\hat{1}'}}:= E_4Y^{(3)}_{\bm{\hat{1}'}}\,,\\
&\dots \,.
\end{aligned}
\end{equation}
Interestingly, there are no VVMFs at weight $k=1$ for the finite modular group $2D_3$.
Alternatively, all modular form multiplets of weights greater than $3$ mentioned above can be obtained through tensor products from the modular form multiplets of weights $2$ and $3$.

As mentioned in section~\ref{sec:2D3group}, all the representations of $2D_3$ coincide with the corresponding representations of $\Gamma'_4\cong S'_4$. Hence, these modular-form multiplets also coincide with those of $\Gamma'_4\cong S'_4$. Therefore, we can construct these modular multiplets also from the modular forms of level $N=4$, as in Refs.~\cite{Novichkov:2020eep,Liu:2020akv}. For example,
\begin{align}
\label{eq:VVMFweight2and5methodB}
Y^{(2)}_{\bm{2}}(\tau)=\begin{pmatrix}
\theta_2^4(2\tau)+\theta_3^4(2\tau) \\
-2\sqrt{3} \theta_2^2(2\tau)\theta_3^2(2\tau)
\end{pmatrix} \,,\quad 
 Y^{(5)}_{\bm{\hat{2}}}(\tau)=\begin{pmatrix}
 \sqrt{3}\left(\theta_2^3(2\tau)\theta_3^7(2\tau)-\theta_2^7(2\tau)\theta_3^3(2\tau) \right)\\
\frac{1}{2}\left(\theta_2(2\tau)\theta_3^9(2\tau)-\theta_2^9(2\tau)\theta_3(2\tau) \right)
\end{pmatrix}\,,
\end{align}
where the Jacobi theta constants $\theta_{2}(2\tau)$ and $\theta_{3}(2\tau)$  can be expressed by the Dedekind eta function $\eta(\tau)$ as
\begin{equation}
 \theta_2(2\tau) ~=~ \frac{2\eta^2(4\tau)}{\eta(2\tau)}
 \qquad\text{and}\qquad
 \theta_3(2\tau) ~=~ \frac{\eta^5(2\tau)}{\eta^2(\tau)\eta^2(4\tau)}\,.
\end{equation}
It is easy to verify from the associated $q$-expansions that the modular multiplets in Eqs.~\eqref{eq:MinimalWeightVVMF2D3} and~\eqref{eq:VVMFweight2and5methodB} indeed coincide.

Notice that the modular-form doublets can always be written as 
\begin{equation}
\begin{aligned}
Y^{(k_Y)}_{\bm{2}}      &~=~ y_1(\tau) Y^{(2)}_{\bm{2}} + y_2(\tau) D_2Y^{(2)}_{\bm{2}}\,, \\
Y^{(k_Y)}_{\bm{\hat{2}}}&~=~ z_1(\tau) Y^{(5)}_{\bm{\hat{2}}} + z_2(\tau) D_2Y^{(2)}_{\bm{\hat{2}}}\,,
\end{aligned}
\end{equation}
where $y_1(\tau)$ and $y_2(\tau)$ are polynomials of $E_4(\tau), E_6(\tau)$, and they are scalar modular forms of $\mathrm{SL}(2,\mathbb{Z})$ with weights $k_Y-2$, $k_Y-4$, respectively. Analogously, $z_1(\tau)$ and $z_2(\tau)$ are modular forms of $\mathrm{SL}(2,\mathbb{Z})$ with weights $k_Y-5$, $k_Y-7$ respectively. The corresponding functions are vanishing if the modular forms at some of these weights do not exist.

%% file: classification.tex
In the formalism of $\mathcal{N}=1$ SUGRA, a model is defined by its K\"ahler potential $\mathcal{K}$ and its superpotential $\mathcal{W}$ (together with a gauge kinetic function, which is unimportant here), where $\Phi_I$ and $\tau$ denote respectively the matter fields and the modulus of the theory. Inspired by string models, $\tau$ may arise from the description of a $\mathbb{T}^2/\Z{N}$ orbifold compactification of two extra dimensions~\cite{Nilles:2020gvu}. In such models, the modular covariant K\"ahler potential in Planck units ($M_\mathrm{Pl}=1$) is given at tree level by~\cite{Dixon:1989fj}
\begin{equation}
\label{eq:kahlerpotential}
\mathcal{K} (\Phi_I , \bar{\Phi}_I ; \tau , \bar{\tau}) ~\supset~ 
-\log{( -\I \tau +\I \bar{\tau})} + \sum_I (-\I \tau +\I \bar{\tau})^{-k_I} |\Phi_I|^2\,.
\end{equation}
Meanwhile, the superpotential is a holomorphic function of the matter fields and the modulus, which can be expressed in power series of the matter fields as
\begin{equation}
\label{eq:Superpotential}
\mathcal{W}(\Phi_I,\tau)~\supset~\sum Y^{(k_Y)}_{\rep{r}_Y;I_1 \dots I_n}(\tau)\ \Phi_{I_1} \cdots \Phi_{I_n}\,,
\end{equation}
where $Y^{(k_Y)}_{\rep{r}_Y;I_1 \dots I_n}(\tau)$ is a VVMF of weight $k_Y$ transforming in the $r_Y$-dimensional representation of a generalized finite modular group $G$.
In the superpotential~\eqref{eq:Superpotential}, just to simplify the notation, we have omitted the numeric (modular-invariant) coefficients of each term in the sum. It is customary to assume that such numeric coefficients are real, which can be the result of imposing an additional \CP or \CP-like symmetry. 

Modular invariance of the theory imposes restrictions on the representations and modular weights of modular forms and matter fields. As also happens naturally  in models arising from string compactifications, we assume that matter fields transform as
\begin{equation}
\label{eq:MatterFieldsTransformation}
 \Phi_I ~\stackrel{\gamma}{\longrightarrow}~ 
   (c\tau+d)^{-k_I} \rho_{\rep{r}_I}(\gamma)\Phi_I\,,\qquad 
 \gamma~=~\begin{pmatrix}
   a & b \\c & d
 \end{pmatrix}\in \Gamma\,,
\end{equation}
where matter fields are proposed to build $r_I$-dimensional irreducible representations of $G$ and to carry modular weights, which are known to be rationals in string models, i.e.~$k_I\in\mathbbm{Q}$, see e.g.~\cite{Ibanez:1992hc,Ferrara:1989qb,Araki:2008ek,Olguin-Trejo:2017zav}. Demanding modular invariance amounts first to imposing
\begin{equation}
\label{eq:CondRepresentations}
  \rep{r}_Y \otimes \rep{r}_{I_1} \otimes \cdots \otimes \rep{r}_{I_n} ~\stackrel!\supset~ \rep{1}\,.
\end{equation}
A second constraint arises from the transformation of the K\"ahler potential. We note that, due to the form of the $\Phi_I$-independent part of Eq.~\eqref{eq:kahlerpotential}, $\mathcal K$ transforms non-trivially under $\gamma\in\Gamma$ as
\begin{equation}
\label{eq:KahlerTrafoOnK}
 \mathcal K~\stackrel{\gamma}{\longrightarrow}~ \mathcal K + f(\tau) + \overline{f(\tau)}
 \qquad\text{with}\qquad
  f(\tau)~:=~\log(c\tau + d)\,. 
\end{equation}
Hence, noting that a K\"ahler transformation~\cite[\S23]{Wess:1992cp} can remove the extra terms in Eq.~\eqref{eq:KahlerTrafoOnK} at the expense of altering the superpotential by $\mathcal W\to \e^{f}\mathcal W=(c\tau+d)\mathcal W$, we conclude that $\mathcal W$ must transform under $\gamma\in\Gamma$ as $\mathcal W \stackrel{\gamma}{\to}(c\tau+d)^{-1}\mathcal W$ to cancel the factor $\e^f$, i.e.~it must have modular weight $k_{\mathcal W} =-1$. Consequently, we are left with the condition
\begin{equation}
\label{eq:fieldsmodularweights}
  k_{Y}-k_{I_1}-\dots-k_{I_n} ~\stackrel!=~ -1\,.
\end{equation}

These general features of modular flavor models can be used to build models based on our particular modular flavor symmetry $G=2D_3$, provided a number of top-down-inspired working assumptions:
\begin{itemize}
\item {\bf\boldmath $2+1$-family structure.\unboldmath} Since $2D_3$ supports only singlet and doublet irreducible representations (see section~\ref{sec:2D3group}), we adopt the natural $2+1$-family structure of the group, where two of the generations (most likely the lighter families) build any of the two doublets whereas the third generation transforms as any of the four singlets of $2D_3$. This implies that three (MSSM) fields with the same gauge quantum numbers build a {\it reducible} triplet that can be written as $\psi = (\psi_D,\psi_3)$, with $\psi_D$ transforming as $\bf2$ or $\bf{\hat2}$, and $\psi_3$ as any of the $2D_3$ singlets $\{\bf1,\bf{1'},\bf{\hat1},\bf{\hat1'}\}$.

\item {\bf Singlet representations for Higgs fields.} For simplicity, our SUGRA models do not exhibit an extended Higgs sector, hence up and down-type Higgs fields are (trivial or non-trivial) singlets of $2D_3$.

\item {\bf\boldmath Modular forms with weight $2\leq k_Y\leq7$.\unboldmath} The higher the modular weight, the more linearly VVMFs there are, which introduces too many parameters into the model. In the interest of simplicity, we restrict ourselves to the Yukawa couplings $Y^{(k_Y)}_{\rep{r}_Y}(\tau)$ given in Eqs.~\eqref{eq:MF-multiplets-2D3}.~\footnote{One must remember that there is also a trivial modular form at weight 0 that can be taken as an arbitrary constant (usually set to 1 for convenience). Beyond this, one could include also combinations of the Klein's $j$-invariant or (modular meromorphic) $j(\tau)$ function~\eqref{eq:j-Delta}. This non-trivial $\tau$-dependence will be studied elsewhere.}

\item {\bf Fractional modular weights for matter generations.} $2D_3$ arises from $\mathbb{T}^2/\Z4$ string orbifolds and the modular weights of matter fields take the form $k_I=\nicefrac{n}{4}$, $n\in\Z{}$,~\cite{Baur:2023}. In addition, as we must satisfy the condition~\eqref{eq:fieldsmodularweights} and $k_Y\leq7$, we assume that 
\begin{equation}
\label{eq:WeightsMatterFields}
 k_I ~\in~ \left\{-4,\tfrac{-15}{4},\tfrac{-7}{2},\dots,\tfrac{7}{2},\tfrac{15}{4},4\right\}\,.
\end{equation}

\item {\bf Vanishing modular weights for Higgs fields.} In the top-down constructions we inspire our study, Higgs fields have modular weights $0,1$, see e.g.~\cite{Parameswaran:2010ec,Baur:2022hma}. Thus, we choose that $k_I=0$ for the Higgs fields of our models.

\end{itemize}  

These properties allow us to draw some general remarks. First, assuming $k_{H_u}=k_{H_d}=0$ and $k_\mathcal{W}=-1$, see Eq.~\eqref{eq:fieldsmodularweights}, implies that the $\mu$-term cannot be generated at the fundamental scale of the theory. Interestingly, this is a key ingredient of the Giudice-Masiero mechanism~\cite{Giudice:1988yz}. In this scenario, a neutral spurion field $X$ has a VEV $\vev{X}$ that breaks supersymmetry spontaneously and induces a hierarchically small $\mu$-term. In our case, the spurion field can be a gauge and flavor singlet with vanishing modular weight. This precise situation happens in the top-down constructions that inspire our assumptions, as studied in~\cite{Chen:2012jg} in terms of $R$ instead of flavor symmetries. Note that one can establish an interesting relation between $R$-charges and modular weights in the top-down approach, see e.g.~\cite[eq.~(105)]{Nilles:2020gvu}. From a bottom-up perspective, since $\mathcal W$ carries nonzero modular weight, modular symmetries are naturally $R$-symmetries. If the spurion contribution to the $\mu$-term was too small, one can additionally consider the inclusion of hierarchically suppressed contributions \`a la Frogatt-Nielsen due to the possible existence of gauge singlets (a hidden sector) with some nontrivial modular weights, as shown in explicit string models~\cite{Kappl:2008ie,Ramos-Sanchez:2010wxr}. Even though the goal of this work is the flavor puzzle and thus the details shall not be worked out here, it is interesting to observe that our scenario can solve more than one problem at once.

As a second remark, as usual in bottom-up models, we must make a couple of assumptions on the K\"ahler potential due to two potential challenges: the existence of non-canonical contributions~\cite{Chen:2019ewa} and the possibility of large corrections due to the signs of the modular weights~\cite{Kobayashi:2023qzt}. Inspired by string-derived models, we shall assume that both issues can be controlled. The former can be controlled by the natural appearance of traditional flavor symmetries, building a so-called eclectic flavor group~\cite{Nilles:2020nnc} that restricts $\mathcal K$ to its canonical form~\cite{Nilles:2020kgo}. The latter challenge can be controlled by considering all other possible features appearing in full string-derived models, which do not only exhibit positive and negative modular weights~\cite{Ibanez:1992hc,Olguin-Trejo:2017zav,Nilles:2020kgo,Baur:2022hma,Kobayashi:2023qzt}, but also extra moduli that can cancel out the dangerous corrections~\cite{Kobayashi:2023qzt}.

With these elements, we are ready to classify all mass textures that arise in models based on $2D_3$ as modular flavor symmetry and the mentioned priors. In the following, we classify separately all Dirac and Majorana mass textures for generic fermions. These structures will be used in the generation of quark and lepton masses in section~\ref{sec:BenchmarkModels}.

%%%%%%%%%%%%%%%%%%%%%%%%%%%%%%%%%%%%
\subsection{Dirac mass textures} 
\label{sec:DiracYukawa} 

Allowing for right-handed neutrinos, as we shall do, up and down quarks as well as charged leptons and neutrinos accept Dirac masses. Thus, it is necessary that we know all possible Dirac mass textures that can arise in models with a $2+1$-family structure. 

In the superpotential~\eqref{eq:Superpotential}, let us denote as $\psi = (\psi_D,\psi_3)$ the reducible flavor triplet corresponding to left-handed fermion $\SU2_L$ doublets and as $\psi^c = (\psi^c_D,\psi^c_3)$ the reducible flavor triplet associated with right-handed fermion $\SU2_L$ singlets. Further, let $\rep{s}$ and $\rep{\tilde{s}}$ denote, beyond $\rep{s},\rep{\tilde{s}}=\rep1$, any of the two pairs of $2D_3$ singlets that deliver the trivial singlet, $\rep{s}\otimes\rep{\tilde{s}}=\rep{1}$, according to the product rules in Eq.~\eqref{eq:SingletProductRules}.
In these terms, the resulting superpotential terms with the Higgs field $H_{u/d}$ transforming as a $\rep{\tilde{s}}$ singlet, takes the form
\begin{align}
\nonumber
\mathcal{W}&\supset\left[\alpha\left(Y^{(k_{DD})}_{DD}\psi^c_D \psi_D\right)_\mathbf{s} 
+ \beta\left(Y^{(k_{D3})}_{D3}\psi^c_D \psi_3\right)_\mathbf{s} 
+ \gamma\left(Y^{(k_{3D})}_{3D}\psi^c_3 \psi_D\right)_\mathbf{s} 
+ \delta\left(Y^{(k_{33})}_{33}\psi^c_3 \psi_3\right)_\mathbf{s}\right] \left(H_{u/d}\right)_{\rep{\tilde{s}}}\,\\
\label{eq:DiracSuperpotential}
&=:~\mathcal{W}_{DD}+\mathcal{W}_{D3}+\mathcal{W}_{3D} + \mathcal{W}_{33}\,,
\end{align}
where $\alpha,\beta,\gamma$ and $\delta$ are some real numeric coefficients. Further, $Y^{(k_{DD})}_{DD}, Y^{(k_{D3})}_{D3}, Y^{(k_{3D})}_{3D}$ and $Y^{(k_{33})}_{33}$ are general VVMFs with modular weights
$k_{DD}:=k_{\psi^c_D}+k_{\psi_D}-1$,
$k_{D3}:=k_{\psi^c_D}+k_{\psi_3}-1$, 
$k_{3D}:=k_{\psi^c_3}+k_{\psi_D}-1$ and 
$k_{33}:=k_{\psi^c_3}+k_{\psi_3}-1$, respectively. 
(Recall that we assume $k_{H_{u/d}}=0$.) The subindex $\rep{s}$ denotes the combination of the components of the fields in parentheses associated with the singlet $\rep{s}$ that results from tensoring the field representations.

The explicit forms of the superpotential components in Eq.~\eqref{eq:DiracSuperpotential} depend on the specific representations and weights of the matter fields. Evidently, the four components of the Dirac superpotential arise from the $2+1$-family structure, and this also implies that the Dirac mass matrix can be divided into four blocks as
\begin{equation}
\label{eq:MassMatrixInBlocks}
M_{\psi}~=~\left(
\begin{array}{c|c}
~M_{DD}^{\phantom{A^A}}~ & M_{D3}\\
\hline
~M_{3D}~ & M_{33}
\end{array}\right)v_{u/d}\,,
\end{equation}
where $v_{u/d}$ denotes the VEV of the Higgs field $H_{u/d}$, and $M_{DD}$, $M_{D3}$, $M_{3D}$ and $M_{33}$ are $2\x2$, $2\x1$, $1\x2$ and $1\x1$ sub-matrices, respectively. Here, $M_\psi$ is given in the right-left basis, such that $\psi^{c}_i(M_\psi)_{ij}\psi_j$ is a superpotential mass term. 

We can now classify all possible Dirac-mass blocks in Eq.~\eqref{eq:MassMatrixInBlocks} arising from all representation configurations $(\psi_a^c,\psi_b,H_{u/d})$ of the fields. Here the indices $a,b$ can be the doublet label $D$ or the singlet label $3$. For example, for $a=D$ and $b=D$ we can take the configuration $(\psi_a^c,\psi_b,H_{u/d})=(\rep2,\bf{\hat2},\bf{1'})$. In this case, we obtain the flavor-invariant 2\x2 block $M_{DD}$ that results from the product $\rep{r_Y}\otimes\rep2\otimes\bf{\hat2}\otimes\bf{1'}$, i.e.~the one associated with $(\rep{r_Y}\otimes\rep2\otimes\bf{\hat2})_{\bf{1'}}\otimes\bf{1'}=\rep1$. Note that, in this case, there are three choices of $\rep{r_Y}$ leading to a $\bf{1'}$ in the parentheses, $\rep{r_Y}\in\{\bf{\hat1},\bf{\hat1'},\bf{\hat2}\}$. We can take care of this ambiguity by setting three different $\alpha_i$, $i=1,2,3$, coefficients. Of course, not all of the resulting terms are relevant in all models. The relevant contributions can only be obtained once the modular weights $k_{DD},k_{\psi^c_D},k_{\psi_D}$ of $Y_{DD}^{(k_{DD})},\psi^c_D,\psi_D$ are identified. For a given weight configuration satisfying Eq.~\eqref{eq:fieldsmodularweights}, there is at most one $\rep{r_Y}$ that also complies with Eq.~\eqref{eq:CondRepresentations}. Thus, at most one $\alpha_i$ is non-vanishing. 

The full classification of all possible Dirac mass blocks is given in Table~\ref{tab:blocks_Dirac} of Appendix~\ref{app:classification_Dirac}. In our illustrative case, the resulting $M_{DD}$ block is labeled as $M_{DD_4}$. Our table includes all 12 independent blocks, four of each independent type $M_{DD}, M_{3D}, M_{D3}$. Once chosen the representation configuration for these blocks, the $M_{33}$ block is automatically defined. 

The full set of possible mass textures is obtained by combining the classified blocks. There is a total of $4\x4\x4=64$ possible Dirac mass textures $M_\psi$, which are presented in Table~\ref{tab:DiracModels} of our Appendix~\ref{app:classification_Dirac}.

The final task to arrive at a model of masses for quarks or leptons is to set the possible modular weights of matter fields $\psi^c$ and $\psi$, which define the modular weights $k_Y\in\{k_{DD},k_{3D},k_{D3},k_{33}\}$ of the VVMFs $Y^{(k_Y)}_{\rep{r_Y}}$ appearing in the mass textures of Table~\ref{tab:DiracModels}. In our survey of models, we consider combinations of the modular weights in Eq.~\eqref{eq:WeightsMatterFields} for the fields in the Yukawa couplings. Combining those weights, we arrive at 158,208 full $3\x3$ Dirac mass textures, but fortunately out of them just 12,316 matrices deliver three non-zero masses. Many of these mass matrices have identical structure; we find that only 1,412 Dirac mass textures with full rank are also inequivalent. We shall use this final set for our scans in section~\ref{sec:BenchmarkModels}.

%%%%%%%%%%%%%%%%%%%%%%%%%%%%%%%%%%%%
\subsection{Majorana mass textures} 
\label{sec:MajoranaYukawa} 
 
It is known that generating the small masses of the three observed neutrino may minimally require, in addition to the Dirac terms that can arise from Eq.~\eqref{eq:DiracSuperpotential}, introducing Majorana terms. In the minimal scenarios of neutrino masses, (symmetric) Majorana mass matrices can arise either from the mass terms for right-handed neutrinos in the type-I seesaw mechanism, or from the so-called Weinberg operator.

If neutrino masses originate from a type-I seesaw mechanism, the superpotential terms corresponding to Majorana textures can be in general written as
\begin{equation}
	\label{eq:MajoranaSuperpotential}
	\mathcal{W}~\supset~\left[\alpha\left(Y^{(k_{DD})}_{DD}\psi^c_D \psi^c_D\right)_{\rep1}
	+ \beta\left(Y^{(k_{D3})}_{D3}\left(\psi^c_D \psi^c_3 + \psi^c_3 \psi^c_D\right)\right)_{\rep1}
	+ \gamma\left(Y^{(k_{33})}_{33}\psi^c_3 \psi^c_3\right)_{\rep1}\ \right] \Lambda
\end{equation}
for a $2+1$-family-structured triplet associated with right-handed fermion $\SU2_L$ singlets. Here, $\alpha,\beta,\gamma$ are some real coefficients. For seesaw neutrino masses, $\Lambda$ denotes the seesaw scale, and the triplet $\psi^c$ corresponds to the multiplet of heavy right-handed neutrino superfields, $(N^c_D,N^c_3)$. Further, from the condition~\eqref{eq:fieldsmodularweights} we see that the modular weights of the various different VVMFs $Y^{(k_{DD})}_{DD}$, $Y^{(k_{D3})}_{D3}$ and $Y^{(k_{33})}_{33}$ must be given by $k_{DD}:=2k_{N^c_D}-1$, $k_{D3}:=k_{N^c_D}+k_{N^c_3}-1$ and $k_{33}:=2k_{N^c_3}-1$.

On the other hand, if neutrino masses arise from the Weinberg operator $LH_uLH_u$, the general structure of the required mass textures reads
\begin{equation}
\begin{aligned}
\label{eq:MajoranaSuperpotential_2}
\mathcal{W}\supset\frac{1}{\Lambda} \left[\alpha\left(Y^{(k_{DD})}_{DD}\psi_D \psi_D\right)_{\rep{s}} \right.
&+ \beta\left(Y^{(k_{D3})}_{D3}\left(\psi_D \psi_3 + \psi_3 \psi_D\right)\right)_{\rep{s}} \\
&\hspace{2.1cm}\left.+ \gamma\left(Y^{(k_{33})}_{33}\psi_3 \psi_3\right)_{\rep{s}}\ \right]{\left(H_{u}H_{u}\right)}_\mathbf{\tilde{s}}\,,
\end{aligned}
\end{equation}                        
where $\Lambda$ denotes the SM lepton-number violation scale, and $\mathbf{s}\otimes\mathbf{\tilde{s}}=\rep{1}$ with $\rep{s}=\rep{\tilde{s}}$ being either $\rep1$ or $\rep{1'}$ according to Eq.~\eqref{eq:SingletProductRules}. As anticipated, the reducible triplet $\psi$ must be the multiplet $(L_D,L_3)$ containing the superfields associated with lepton $\SU2_L$ doublets. In this case, the modular weights of the different VVMFs $Y^{(k_{DD})}_{DD}$, $Y^{(k_{D3})}_{D3}$ and $Y^{(k_{33})}_{33}$ are respectively given by $k_{DD}:=2k_{L_D}-1$, $k_{D3}:=k_{L_D}+k_{L_3}-1$, $k_{33}:=2k_{L_3}-1$, and $k_{H_u}=0$ by choice.

As for the Dirac case in section~\ref{sec:DiracYukawa}, we can perform an analogous procedure splitting the Majorana mass texture $M_\psi$ in block structure, where $v_{u/d}$ should be replaced by either $\Lambda$ or $v_u^2/\Lambda$, depending on the mechanism that yields the Majorana masses. Then, we must combine them to build the $3\x3$ mass matrices. Interestingly, we find only the eight different cases listed in Table~\ref{tab:MajoranaModels} of our Appendix~\ref{app:classification_Majorana}, where the modular weights of the $2D_3$ VVMFs must still be fixed. After setting modular weights for matter fields from the set in Eq.~\eqref{eq:WeightsMatterFields}, we find 312 cases, out of which there are only 52 inequivalent Majorana mass textures that deliver three non-vanishing masses.

%% file: models.tex
Based on the general classification of fermion mass matrices in the previous section, all possible quark or lepton mass models can be constructed, building a huge set of models. Among them we must identify only the ones that best fit observations with the least number of parameters. In fact, since different choices of representations and weights for matter fields deliver the same textures and these are the ones responsible for flavor predictions, we must only identify the best textures for quark and lepton masses.   

Whether a mass model is realistic or not can be assessed by the conventional $\chi^2$ function
\begin{equation}
\label{eq:chi2}
  \chi^2(x) ~=~ \sum_i \left( \frac{\mu_{i,\text{exp}}-\mu_{i,\text{model}}(x)}{\sigma_i} \right)^2\,,
\end{equation}
where the vector $x$ represents the input parameters of the model, $\mu_{i,\text{model}}$ are the model predictions for flavor observables, which include fermion mass ratios, mixing angles and \CP violation phases, $\mu_{i,\text{exp}}$ and $\sigma_i$ are the corresponding experimental central values and $1\sigma$ errors. The observable data we use for leptons and quarks are summarized in Table~\ref{tab:experimental_data}. Because the leptonic Dirac \CP phase $\delta^{\ell}_{\CP}$ has not been accurately measured, it will not be included in $\chi^2$ function.

\begin{table}[b!]
\centering
\resizebox{0.85\textwidth}{!}{
\begin{tabular}{| c | c || c | c|} \hline \hline
\multicolumn{2}{|c||}{Quark sector}               &\multicolumn{2}{c|}{Lepton sector} \\ \hline
Observables & Central value and $1\sigma$ error   &  Observables                           & Central value and $1\sigma$ error\\ \hline
$m_u/m_c$ & $(1.9286\pm{0.6017})\times 10^{-3}$   & $m_e/m_\mu $                           & $0.00474\pm{0.00004}$ \\
$m_c/m_t$ & $(2.8213\pm{0.1195})\times 10^{-3}$   & $m_\mu/m_\tau$                         & $0.0586\pm{0.00047}$ \\
$m_d/m_s$ & $(5.0523\pm{0.6191})\times 10^{-2}$   & $\Delta m_{21}^2 / 10^{-5}\text{eV}^2$ & $7.41^{+0.21}_{-0.20}$ \\
$m_s/m_b$ & $(1.8241\pm{0.1005})\times 10^{-2}$   & $\Delta m_{31}^2 / 10^{-3}\text{eV}^2$ & $2.511^{+0.028}_{-0.027}$  \\ \hline
$\delta^q_{\CP}/{}^\circ$ & $69.2133\pm{3.1146}$  & $\delta^{\ell}_{\CP}/\pi$              & $1.0944^{+0.2333}_{-0.1389}$ \\
$\theta^q_{12}$           & $0.22736\pm{0.00073}$ & $\sin^2\theta^{\ell}_{12}$             & $0.303^{+0.012}_{-0.011}$  \\
$\theta^q_ {13}$          & $0.00349\pm{0.00013}$ & $\sin^2\theta^{\ell}_{13}$             & $0.02203^{+0.00056}_{-0.00059}$  \\
$\theta^q_{23}$           & $0.04015\pm{0.00064}$ & $\sin^2\theta^{\ell}_{23}$             & $0.572^{+0.018}_{-0.023}$  \\ 
\hline \hline
\end{tabular}}
\caption{Experimental central values and $1\sigma$ errors of the mass ratios, mixing angles and \CP-violation phases for quark and lepton sectors. The data of charged lepton mass ratios, quark mass ratios and quark mixing parameters are taken from Ref.~\cite{Antusch:2013jca} with $M_\mathrm{SUSY}=10$\,TeV and $\tan\beta=10, \bar{\eta}_b=0$. The lepton mixing parameters are taken from NuFIT 5.2 (2022)~\cite{Esteban:2020cvm} for normal ordered neutrino masses,
without including Super-Kamiokande atmospheric data. 
\label{tab:experimental_data}
}
\end{table}

The values of the input parameters at the minimum of the $\chi^2$ function of a model can be regarded as their best-fit values. A model can be considered as a phenomenologically viable benchmark model if the observables predicted at the best-fit values fall within the $3\sigma$ range of the experimental data. Recall that, in order to reduce the number of free parameters, we have assumed an underlying  \CP symmetry. Further, as shown in Table~\ref{tab:2D3reps}, the representation matrices of the generators $S$ and $T$ are taken to be symmetric, and all Clebsch-Gordan coefficients are real. Consequently, all numeric coefficients in the superpotential, such as $\alpha_i,\beta_i, \gamma_i$ and so forth, can be constrained to be real numbers. In this way, any complex phases or \CP violation within the theory are exclusively attributed to the \CP-violating modulus VEV $\braket{\tau}$~\cite{Novichkov:2019sqv,Ishiguro:2020nuf,Ding:2021iqp}. With these priors, we can now build some benchmark models for quarks and leptons.

%%%%%%%%%%%%%%%%%%%%%%%%%%%%%%%%%%%%
\subsection{Quark model}
\label{sec:quarkmodel}

We were unable to identify a fully phenomenologically viable model of quarks with less than 10 parameters. We find some models with 9 parameters that are mostly compatible with observations, excepting for one mass ratio that is slightly out of the $3\sigma$ range. Fortunately, there are many fully viable 10-parameter quark models. Therefore, in this section we give an example of a semi-viable model with 9 parameters and a viable model with 10 parameters.	

The modular weights and representations defining our 9-parameter quark-mass model are
\begin{equation}
\begin{aligned}
&d^c_D\oplus d^c_3 \sim \mathbf{\hat{2}}\oplus \rep{1}\,,\quad u^c_D\oplus u^c_3 \sim \mathbf{\hat{2}}\oplus \mathbf{\hat{1}}\,,\quad Q_D\oplus Q_3 \sim \mathbf{\hat{2}}\oplus \rep{1} \,,\quad H_d \sim \mathbf{\hat{1}'}\,,~~H_u \sim \mathbf{1}\,.\\
&k_{d^c_D}=3\,,\quad k_{d^c_3}=k_{u^c_3}=2\,,\quad k_{Q_D}=1\,,\quad k_{Q_3}=k_{u^c_D}=4\,, \quad k_{H_d}=k_{H_u}=0\,.
\end{aligned}
\end{equation}
This leads to the superpotential terms $\mathcal{W}\supset \mathcal{W}_d+\mathcal{W}_u$ for down and up quark masses, where
\begin{align}
\mathcal{W}_{d}&= \left[\alpha^d_3 \left(d_{D}^{c} Q_D Y^{(3)}_{\mathbf{\hat{1}'}} \right)_{\mathbf{\hat{1}}}+ \beta^d \left(d_{D}^{c} Q_3 Y^{(6)}_{\mathbf{2}}\right)_{\mathbf{\hat{1}}} +\gamma^d \left(d^c_{3} Q_D Y^{(2)}_{\mathbf{2}}\right)_{\mathbf{\hat{1}}} \right] H_d\,,\\
\nonumber
\mathcal{W}_{u}&=\left[\alpha_1^u \left(u^c_{D} Q_D Y^{(4)}_{\mathbf{2}}\right)_{\mathbf{1}}+ \alpha_2^u \left(u^c_{D} Q_D Y^{(4)}_{\mathbf{1}}\right)_{\mathbf{1}} +\beta^u \left(u_{D}^{c} Q_3 Y^{(7)}_{\mathbf{\hat{2}}}\right)_{\mathbf{1}}+\gamma^u \left(u^c_{3} Q_D Y^{(2)}_{\mathbf{2}}\right)_{\mathbf{1}}\right] H_{u}\,.
\end{align}
We see that this model corresponds to $\mathcal{D}_{17}$ and $\mathcal{D}_{42}$ mass textures of Table~\ref{tab:DiracModels} for down and up quarks, respectively, which are given for the chosen weights by
\begin{subequations}
\label{eq:QuarkModel1MassMtrices}
\begin{align} 
\label{eq:QuarkModel1MassMtricesMd}
M_{d}&= \begin{pmatrix}
	\alpha^d_3 Y^{(3)}_{\mathbf{\hat{1}'}} & 0 & \beta^d Y^{(6)}_{\mathbf{2},1} \\
	0 & \alpha^d_3 Y^{(3)}_{\mathbf{\hat{1}'}} & \beta^d Y^{(6)}_{\mathbf{2},2}\\
	\gamma^d Y^{(2)}_{\mathbf{2},1} & \gamma^d Y^{(2)}_{\mathbf{2},2}  &  0 \end{pmatrix} v_{d}\,,\\
\label{eq:QuarkModel1MassMtricesMu}
M_u &= \begin{pmatrix}
	\alpha^u_1 Y^{(4)}_{\mathbf{2},2} & \alpha^u_1 Y^{(4)}_{\mathbf{2},1}+\alpha^u_2 Y^{(4)}_{\mathbf{1}}  & \beta^u Y^{(7)}_{\mathbf{\hat{2}},2}  \\
	\alpha^u_1 Y^{(4)}_{\mathbf{2},1}-\alpha^u_2 Y^{(4)}_{\mathbf{1}}  &  -\alpha^u_1 Y^{(4)}_{\mathbf{2},2}  & -\beta^u Y^{(7)}_{\mathbf{\hat{2}},1} \\
	-\gamma^u Y^{(2)}_{\mathbf{2},2}  & \gamma^u Y^{(2)}_{\mathbf{2},1}  & 0 \\
\end{pmatrix}v_u\,.
\end{align}
\end{subequations}
By minimizing the $\chi^2$ function, we identify the best-fit dimensionless parameters
\begin{equation}
\label{eq:Quark9parabf}
\begin{aligned}
\braket{\tau} &=-0.484521+0.891827\I\,,\quad &\beta^d/\alpha_3^d&=-0.264015\,,\quad &\gamma^d/\alpha_3^d&=11.457\,,\\
\alpha_1^u/\gamma^u&=-450.529\,,~&\alpha_2^u/\gamma^u&=123.919\,,~ &\beta^u/\gamma^u&=-0.63101\,,
\end{aligned}
\end{equation}
which deliver the mass ratios and mixing angles given by
\begin{eqnarray}
\nonumber&&m_u/m_c=0.00193121\,,~ m_c/m_t=0.00287386\,,~ m_d/m_s=0.0230934\,,~ m_s/m_b=0.018243\,,\\
&& \theta^{q}_{12}=0.227392\,,\quad \theta^{q}_{13}=0.00349376\,,\quad \theta^{q}_{23}=0.0400987\,,\quad \delta_{\CP}^{q}=70.2738^\circ\,,
\end{eqnarray}
with $\chi^{2}_{\rm min}\approx 19.9$.  Comparing with the observable data of Table~\ref{tab:experimental_data}, we find a great compatibility within $1\sigma$ of three mixing angles, the quark \CP-violation phase and three mass ratios. Only $m_d/m_s$ is smaller than the experimental central value by about $4\sigma$.
Interestingly, the best-fit value of the modulus of this model is close to the critical point $\omega:=\e^{\nicefrac{2\pi \I}{3}}$, and there are no large hierarchies between the input parameters.

For our 10-parameters quark mass model, the quark fields transform under the $2D_3$ modular symmetry as
\begin{align}
\nonumber
&d^c_D\oplus d^c_3 \sim \mathbf{2}\oplus \mathbf{1}\,,\quad u^c_D\oplus u^c_3 \sim \mathbf{\hat{2}}\oplus \mathbf{1}\,,\quad  Q_D\oplus Q_3 \sim \mathbf{\hat{2}}\oplus \mathbf{1} \,,\quad H_d \sim \mathbf{1}\,,~~H_u \sim \mathbf{1}\,.\\
&k_{d^c_D}=k_{Q_D}=k_{u^c_3}=4\,,\quad k_{d^c_3}=0\,,\quad k_{Q_3}=k_{u^c_D}=1\,, \quad k_{H_d}=k_{H_u}=0\,.
\label{eq:10paramsDef}
\end{align}
The corresponding contributions to the modular-invariant superpotential are given by 
\begin{align}
\mathcal{W}_{d}&= \left[\alpha^d_1 \left(d_{D}^{c} Q_D Y^{(7)}_{\mathbf{\hat{2}}} \right)_{\mathbf{1}}+\alpha^d_3 \left(d_{D}^{c} Q_D Y^{(7)}_{\mathbf{\hat{1}'}} \right)_{\mathbf{1}}+ \beta^d \left(d_{D}^{c} Q_3 Y^{(4)}_{\mathbf{2}}\right)_{\mathbf{1}} + \delta^d \left(d_{3}^{c} Q_3\right)_{\mathbf{1}}\right] H_d\,,\\\nonumber
\mathcal{W}_{u}&=\left[\alpha_1^u \left(u^c_{D} Q_D Y^{(4)}_{\mathbf{2}}\right)_{\mathbf{1}}+ \alpha_2^u \left(u^c_{D} Q_D Y^{(4)}_{\mathbf{1}}\right)_{\mathbf{1}} +\gamma^u \left(u^c_{3} Q_D Y^{(7)}_{\mathbf{\hat{2}}}\right)_{\mathbf{1}}+\delta^u \left(u^c_{3} Q_3 Y^{(4)}_{\mathbf{1}}\right)_{\mathbf{1}}\right] H_{u}\,.
\end{align}
The corresponding down and up-quark mass textures are labeled as $\mathcal{D}_{19}$ and $\mathcal{D}_{43}$ in Table~\ref{tab:DiracModels}, respectively. With our chosen weights~\eqref{eq:10paramsDef}, the mass matrices read
\begin{subequations}
\label{eq:QuarkModel2MassMtrices}
\begin{align} 
\label{eq:QuarkModel2MassMtricesMd}
M_{d}&~=~ \begin{pmatrix}
	\alpha^d_1 Y^{(7)}_{\mathbf{\hat{2}},2}+\alpha^d_3 Y^{(7)}_{\mathbf{\hat{1}'}} & \alpha^d_1 Y^{(7)}_{\mathbf{\hat{2}},1} & \beta^d Y^{(4)}_{\mathbf{2},1} \\
	\alpha^d_1 Y^{(7)}_{\mathbf{\hat{2}},1}& -\alpha^d_1 Y^{(7)}_{\mathbf{\hat{2}},2}+\alpha^d_3 Y^{(7)}_{\mathbf{\hat{1}'}} & \beta^d Y^{(4)}_{\mathbf{2},2}\\
	0 & 0 &  \delta^d \end{pmatrix} v_{d}\,,\\
\label{eq:QuarkModel2MassMtricesMu}
M_u &~=~ \begin{pmatrix}
	\alpha^u_1 Y^{(4)}_{\mathbf{2},2} & \alpha^u_1 Y^{(4)}_{\mathbf{2},1}+\alpha^u_2 Y^{(4)}_{\mathbf{1}}  & 0  \\
	\alpha^u_1 Y^{(4)}_{\mathbf{2},1}-\alpha^u_2 Y^{(4)}_{\mathbf{1}}  &  -\alpha^u_1 Y^{(4)}_{\mathbf{2},2}  & 0 \\
	\gamma^u Y^{(7)}_{\mathbf{\hat{2}},2}  & -\gamma^u Y^{(7)}_{\mathbf{\hat{2}},1}  & \delta^u Y^{(4)}_{\mathbf{1}} \\
\end{pmatrix}v_u\,.
\end{align}
\end{subequations}
The best-fit values of the dimensionless input parameters in this model are 
\begin{equation}
\label{eq:Quark10parabf}
\begin{aligned}
\braket{\tau}&=-0.0458127+1.07561\I\,,\quad &\alpha_3^d/\alpha_1^d&=1.16987\,,\quad &\beta^d/\alpha_1^d&=7.00978\,, \\
\delta^d/\alpha_1^d&=0.62122\,,\quad\alpha_1^u/\gamma^u=0.0187396\,,~     &\alpha_2^u/\gamma^u&=0.0186667\,,~ &\delta^u/\gamma^u&=-13.3257\,,
\end{aligned}
\end{equation}
leading to the model's predictions for flavor observables
\begin{eqnarray}
\nonumber&&m_u/m_c=0.00192771\,,~ m_c/m_t=0.00282204\,,~ m_d/m_s=0.0505254\,,~ m_s/m_b=0.0182414\,,\\
&& \theta^{q}_{12}=0.227368\,,\quad \theta^{q}_{13}=0.00349295\,,\quad \theta^{q}_{23}=0.0401456\,,\quad \delta_{\CP}^{q}=69.2212^\circ\,,
\end{eqnarray}
with $\chi^{2}_{\rm min}\approx 0.0002$.  
Comparing with Table~\ref{tab:experimental_data}, these predictions are compatible with the experimental data within the $1\sigma$ interval.
Interestingly, the best-fit value of modulus of this model is close to the critical point $\I$, and there is no large hierarchies between the input parameters.

%%%%%%%%%%%%%%%%%%%%%%%%%%%%%%%%%%%%
\subsection{Dirac neutrino model}
\label{sec:Diracneutino}

Let us study the possibility of neutrino masses arising purely from Yukawa couplings, which requires assuming the existence of extra right-handed neutrinos $N^c:=(N^c_D,N_3^c)$ with a $2+1$-family structure. In this case, we find that at least 8 input parameters are needed to agree with the experimental values of lepton flavor observables. There are about 7,000 non-trivial Dirac neutrino models with 8 parameters, and about 15\% of them are compatible with observations. We present four benchmark examples of Dirac neutrino models here, labeled as D1, D2, D3 and D4. A summary of their weight and representation assignments can be found at the top half of Table~\ref{tab:DiracNeutrinoModels}. It should be noted that the negative weights $k_{N^c}$ assigned to the heavy neutrino fields $N^c$ naturally prohibit the presence of Majorana mass terms in these models.\footnote{On the other hand, the negative $k_{N^c}$ also leads to the large weight $k_L$. We admit this to happen because the weights of the VVMFs that appear in those models are still within our preset range.}  
\begin{table}[h!]
\centering
\resizebox{0.85\textwidth}{!}{
\begin{tabular}{|c|ccccc|cccc|}
\hline\hline
Models &$\rho_{E^c}$ &$\rho_{L}$ &$\rho_{N^c}$ & $\rho_{H_d}$ & $\rho_{H_u}$ &$k_{E^c_D},~k_{E^c_3}$ &$k_{L_D},~k_{L_3}$ &$k_{N^c_D},k_{N^c_3}$ & $k_{H_{u/d}}$ \\ \hline
D1 &$\mathbf{2}\oplus\mathbf{1}$ &$\mathbf{2}\oplus\mathbf{\hat{1}'}$&$\mathbf{2}\oplus\mathbf{\hat{1}'}$ & $\mathbf{1'}$ & $\mathbf{\hat{1}}$ &$-5/2,-5/2$ &$11/2,13/2$ &$-3/2,-1/2$ & $0$ \\ \hline
D2 &$\mathbf{2}\oplus\mathbf{\hat{1}}$ &$\mathbf{\hat{2}}\oplus\mathbf{1'}$&$\mathbf{2}\oplus\mathbf{\hat{1}}$ & $\mathbf{1}$ & $\mathbf{1}$ &$-7/2,-1/2$ &$15/2,17/2$ &$-3/2,-1/2$ & $0$ \\ \hline
D3 &$\mathbf{2}\oplus\mathbf{1}$ &$\mathbf{2}\oplus\mathbf{\hat{1}}$&$\mathbf{2}\oplus\mathbf{1'}$ & $\mathbf{1'}$ & $\mathbf{1}$ &$-1/2,-1/2$ &$7/2,13/2$ &$-1/2,-1/2$ & $0$ \\ \hline
D4 &$\mathbf{2}\oplus\mathbf{\hat{1}'}$ &$\mathbf{2}\oplus\mathbf{\hat{1}}$&$\mathbf{2}\oplus\mathbf{\hat{1}'}$ & $\mathbf{\hat{1}'}$ & $\mathbf{\hat{1}'}$ &$-1/2,-7/2$ &$13/2,11/2$ &$-1/2,-3/2$ & $0$ \\ \hline
\end{tabular}} \\[6pt]
\resizebox{0.85\textwidth}{!}{
\begin{tabular}{|c|c| c|c|c|c|}\hline
Model &D1 &Model &D2 &D3 &D4  \\ \hline
$\text{Re}\langle\tau\rangle$ &-0.0323914 &$\text{Re}\langle\tau\rangle$ & 0.484077 & 0.146315 &0.42963  \\ \hline
$\text{Im}\langle\tau\rangle$ & 1.28408 &$\text{Im}\langle\tau\rangle$ &1.37015 & 0.989238 & 1.7516  \\ \hline
$\delta^{\ell}/\alpha^{\ell}$ &-0.634163 &$\beta^{\ell}/\alpha^{\ell}$ &-379.019  &-1057.33 &-13.1399 \\ \hline
$\gamma^{\ell}/\alpha^{\ell}$ & -17.0629 & $\gamma^{\ell}/\alpha^{\ell}$ &-25.0074 &-3789.24 &-225.982  \\ \hline
$\alpha^{\nu}/\gamma^{\nu}$ &-3.13884 &$\alpha^{\nu}/\gamma^{\nu}$&-6.06662 & -0.763583 &12.6598 \\ \hline
$\beta^{\nu}/\gamma^{\nu}$ &-0.402086 &$\beta^{\nu}/\gamma^{\nu}$& 1.12118 &5.80788 &1.39089 \\ \hline
$\alpha^{\ell}v_d[\MeV]$ &73.4193 &$\alpha^{\ell}v_d [\MeV]$ &3.14488 &0.285387 &5.75080  \\ \hline
$\gamma^{\nu}v_u[\meV]$ & 45.5403 & $\gamma^{\nu}v_u[\meV]$  &40.9040  &30.8341 &32.5682 \\\hline \hline
$m_e/m_\mu$ &0.00473701 &$m_e/m_\mu$ &0.00473696  &0.00473706 &0.00473702 \\ \hline
$m_\mu/m_\tau$ &0.058568 &$m_\mu/m_\tau$ &0.0585677 &0.0585678 &0.0585678  \\ \hline
$\sin^2\theta^{\ell}_{12}$ &0.302989 &$\sin^2\theta^\ell_{12}$ &0.302963  &0.305668 &0.302990  \\ \hline
$\sin^2\theta^{\ell}_{13}$ &0.0220300 &$\sin^2\theta^\ell_{13}$ &0.0220303 &0.0218065 &0.0220301  \\ \hline
$\sin^2\theta^{\ell}_{23}$ &0.571993 &$\sin^2\theta^\ell_{23}$ &0.571986 &0.492503 &0.571995  \\ \hline
$\delta^{\ell}_{\CP}/\pi$ &1.49060 &$\delta^{\ell}_{\CP}/\pi$ &1.68648 &1.00000 &1.30863  \\ \hline
$m_1[\meV]$ &17.5723 &$m_1[\meV]$ &29.8436 &28.9179 &25.6255 \\ \hline
$m_2 [\meV]$ &19.5675 &$m_2[\meV]$ &31.0603 & 30.1720 & 27.0327  \\ \hline
$m_3 [\meV]$ &53.1012 &$m_3[\meV]$ &58.3225  &57.2755 & 56.2816  \\ \hline
$m_{\beta} [\meV]$ &19.6484 &$m_{\beta}[\meV]$ &31.1112 &30.1944 & 27.09129  \\ \hline
$\chi^2_{\text{min}}$ & $10^{-6}$ & $\chi^2_{\text{min}}$ & $10^{-5}$  & 16.09 & $10^{-5}$ \\ \hline\hline
\end{tabular}}
\caption{\label{tab:DiracNeutrinoModels}Summary of the four benchmark Dirac neutrino models based on $2D_3$ modular symmetry. The assignments of the representations and weights of these models are not unique, just chosen to exemplify the potential of our framework. The concrete forms of their respective mass matrices can be found in Eq.~\eqref{eq:DiracModelMassMatrices}. The best-fit values of the input parameters and the corresponding predictions for neutrino mixing angles, and Dirac \CP-violation phase, and the neutrino masses are also included.}
\end{table}

%\newpage
The mass textures for these models can be read directly from Table~\ref{tab:DiracModels}. Specifically, the charged-lepton and neutrino mass matrices of model D1 arise from the textures labeled $(\mathcal{D}_{42}, \mathcal{D}_{17})$, the matrices for model D2 come from $(\mathcal{D}_{22}, \mathcal{D}_{22})$, those for model D3 come from $(\mathcal{D}_{46}, \mathcal{D}_{10})$, and the textures for model D4 arise from  $(\mathcal{D}_{50}, \mathcal{D}_{50})$. Given our chose of weights, the resulting matrices are explicitly given by
\begin{align} 
\label{eq:DiracModelMassMatrices}
\nonumber
\text{D1}:&~~
M_{e} = \begin{pmatrix}
\alpha^{\ell} Y^{(2)}_{\mathbf{2},2} & \alpha^{\ell} Y^{(2)}_{\mathbf{2},1} & 0 \\
\alpha^{\ell} Y^{(2)}_{\mathbf{2},1} & -\alpha^{\ell} Y^{(2)}_{\mathbf{2},2} & 0 \\
-\gamma^{\ell} Y^{(2)}_{\mathbf{2},2} &\gamma^{\ell} Y^{(2)}_{\mathbf{2},1} & \delta^{\ell} Y^{(3)}_{\mathbf{\hat{1}'}}\
\end{pmatrix} v_{d}\,, ~~ M_\nu = 
\begin{pmatrix}
\alpha^\nu Y^{(3)}_{\mathbf{\hat{1}'}} & 0 & \beta^\nu Y^{(4)}_{\mathbf{2},1}  \\
0 &  \alpha^\nu Y^{(3)}_{\mathbf{\hat{1}'}}  &\beta^\nu Y^{(4)}_{\mathbf{2},2} \\
\gamma^\nu Y^{(4)}_{\mathbf{2},1}  & \gamma^\nu Y^{(4)}_{\mathbf{2},2}  & 0 \\
\end{pmatrix}\frac{v_{u}^{2}}{\Lambda}
\,.\\[3pt]
\nonumber
\text{D2}:&~~
M_{e} = \begin{pmatrix}
\alpha^{\ell} Y^{(3)}_{\mathbf{\hat{1}'}} & 0 & -\beta^{\ell} Y^{(4)}_{\mathbf{2},2}  \\
0 &  \alpha^{\ell} Y^{(3)}_{\mathbf{\hat{1}'}}  &\beta^{\ell} Y^{(4)}_{\mathbf{2},1} \\
-\gamma^{\ell} Y^{(6)}_{\mathbf{2},2}  & \gamma^{\ell} Y^{(6)}_{\mathbf{2},1}  & 0 \\
\end{pmatrix} v_{d}\,, ~~ M_\nu = 
\begin{pmatrix}
\alpha^\nu Y^{(5)}_{\mathbf{\hat{2}},2} & \alpha^\nu Y^{(5)}_{\mathbf{\hat{2}},1} & -\beta^\nu Y^{(6)}_{\mathbf{2},2}  \\
\alpha^\nu Y^{(5)}_{\mathbf{\hat{2}},1} &  -\alpha^\nu Y^{(5)}_{\mathbf{\hat{2}},2}  &\beta^\nu Y^{(6)}_{\mathbf{2},1}\\
-\gamma^\nu Y^{(6)}_{\mathbf{2},2}  & \gamma^\nu Y^{(6)}_{\mathbf{2},1}  & 0 \\
\end{pmatrix}\frac{v_{u}^{2}}{\Lambda}
\,.\\[3pt]
\nonumber
\text{D3}:&~~
M_{e} = \begin{pmatrix}
\alpha^{\ell} Y^{(2)}_{\mathbf{2},2} & \alpha^{\ell} Y^{(2)}_{\mathbf{2},1} & \beta^{\ell} Y^{(5)}_{\mathbf{\hat{2}},1}  \\
\alpha^{\ell} Y^{(2)}_{\mathbf{2},1}  &  -\alpha^{\ell} Y^{(2)}_{\mathbf{2},2}   &\beta^{\ell} Y^{(5)}_{\mathbf{\hat{2}},2}  \\
-\gamma^{\ell} Y^{(2)}_{\mathbf{2},2}  & \gamma^{\ell} Y^{(2)}_{\mathbf{2},1}  & 0 \\
\end{pmatrix} v_{d}\,, ~~ M_\nu = 
\begin{pmatrix}
-\alpha^\nu Y^{(2)}_{\mathbf{2},1} & \alpha^\nu Y^{(2)}_{\mathbf{2},2} & \beta^\nu Y^{(5)}_{\mathbf{\hat{2}},2}  \\
\alpha^\nu Y^{(2)}_{\mathbf{2},2} & \alpha^\nu Y^{(2)}_{\mathbf{2},1} & -\beta^\nu Y^{(5)}_{\mathbf{\hat{2}},1} \\
-\gamma^\nu Y^{(2)}_{\mathbf{2},2}  & \gamma^\nu Y^{(2)}_{\mathbf{2},1}  & 0 \\
\end{pmatrix}\frac{v_{u}^{2}}{\Lambda}
\,.\\[3pt]
\nonumber
\text{D4}:&~~
M_{e} = \begin{pmatrix}
-\alpha^{\ell} Y^{(5)}_{\mathbf{\hat{2}},1} & \alpha^{\ell} Y^{(5)}_{\mathbf{\hat{2}},2} & \beta^{\ell} Y^{(4)}_{\mathbf{2},1}  \\
\alpha^{\ell} Y^{(5)}_{\mathbf{\hat{2}},2}  &  \alpha^{\ell} Y^{(5)}_{\mathbf{\hat{2}},1}   &\beta^{\ell} Y^{(4)}_{\mathbf{2},2}  \\
-\gamma^{\ell} Y^{(2)}_{\mathbf{2},2}  & \gamma^{\ell} Y^{(2)}_{\mathbf{2},1}  & 0 \\
\end{pmatrix} v_{d}\,, ~~ M_\nu = 
\begin{pmatrix}
-\alpha^\nu Y^{(5)}_{\mathbf{\hat{2}},1} & \alpha^\nu Y^{(5)}_{\mathbf{\hat{2}},2} & \beta^\nu Y^{(4)}_{\mathbf{2},1}  \\
\alpha^\nu Y^{(5)}_{\mathbf{\hat{2}},2} & \alpha^\nu Y^{(5)}_{\mathbf{\hat{2}},1} & \beta^\nu Y^{(4)}_{\mathbf{2},2}\\
-\gamma^\nu Y^{(4)}_{\mathbf{2},2}  & \gamma^\nu Y^{(4)}_{\mathbf{2},1}  & 0 \\
\end{pmatrix}\frac{v_{u}^{2}}{\Lambda}
\,.\\
\end{align}

Interestingly, in all models the predicted sum of neutrino masses is near the upper bound of the Planck Collaboration result~\cite{Planck:2018vyg}, $\sum_{i}m_{i}<120~\text{meV}$. For models D1, D2 and D4, their predicted Dirac \CP-violation phases $\delta_{\CP}$ are near $1.5\pi$, while model D3 predicts a trivial Dirac \CP phase because the associated best-fit modulus value falls right on the \CP-conserving boundary of the fundamental domain, $|\braket{\tau}|=1$.

%%%%%%%%%%%%%%%%%%%%%%%%%%%%%%%%%%%%
\subsection{Majorana neutrino model for type-I seesaw mechanism}
\label{sec:Seesawneutino}

Let us now explore the generation of neutrino masses by means of the type-I seesaw mechanism. In this case, we find that we need at least 9 parameters to arrive at a realistic model. Out of approximately 6,000 non-trivial seesaw neutrino models with 9 parameters, about 25\% of them are compatible with observations. We present a benchmark model here, in which the modular weights and representations of the matter fields are given by
\begin{align}
\nonumber
&E^c_D\oplus E^c_3 \sim \mathbf{2}\oplus \mathbf{\hat{1}}\,,\quad L_D\oplus L_3 \sim \mathbf{2}\oplus \mathbf{\hat{1}'} \,,\quad N^c_D\oplus N^c_3 \sim \mathbf{2}\oplus \mathbf{1} \,,\quad H_d \sim \mathbf{\hat{1}'}\,,~~H_u \sim \mathbf{1}\,.\\
&k_{E^c_D}=5/2\,,\quad k_{E^c_3}=k_{L_D}=k_{N^c_D}=k_{N^c_3}=3/2\,,\quad k_{L_3}=9/2\,, \quad k_{H_d}=k_{H_u}=0\,.\label{eq:seesawDef}
\end{align}
These choices lead to the $2D_3$ modular invariant superpotential contributions
\begin{equation}
\begin{aligned}
\mathcal{W}_{e}&~=~ \left[\alpha^{\ell} \left(E_{D}^{c} L_D Y^{(3)}_{\mathbf{\hat{1}'}}\right)_{\mathbf{\hat{1}}}+ \beta^{\ell} \left(E_{D}^{c} L_3 Y^{(6)}_{\mathbf{2}}\right)_{\mathbf{\hat{1}}} +\gamma^{\ell}\left(E_{3}^{c} L_D Y^{(2)}_{\mathbf{2}}\right)_{\mathbf{\hat{1}}} \right] H_d\,,\\
\mathcal{W}_{\nu}&~=~ \left[\alpha^\nu \left(N^c_{D} L_D Y^{(2)}_{\mathbf{2}}\right)_{\mathbf{1}}+ \beta^\nu \left(N^c_{D} L_3 Y^{(5)}_{\mathbf{\hat{2}}}\right)_{\mathbf{1}} +\gamma^\nu \left(N^c_{3} L_D Y^{(2)}_{\mathbf{2}}\right)_{\mathbf{1}}\right] H_{u} \\
&~+\left[\alpha^N \left(N^c_{D} N^c_D Y^{(2)}_{\mathbf{2}}\right)_{\mathbf{1}}+ \beta^N \left(N^c_{D} N^c_3 Y^{(2)}_{\mathbf{2}}\right)_{\mathbf{1}}+ \beta^N \left(N^c_3 N^c_{D} Y^{(2)}_{\mathbf{2}}\right)_{\mathbf{1}} \right]\Lambda\,.
\end{aligned}
\end{equation}
Note that $\mathcal W_\nu$ includes Yukawa terms and Majorana contributions to the masses, corresponding to two different texture contributions from Tables~\ref{tab:DiracModels} and~\ref{tab:MajoranaModels}. From the model definition~\eqref{eq:seesawDef}, we see that in the notation of our Appendices~\ref{app:classification_Dirac} and~\ref{app:classification_Majorana}, the Dirac mass texture for charged leptons is $\mathcal{D}_{53}$, while the two textures for neutrino masses are labeled $\mathcal{D}_{13}$ and $\mathcal{N}_{1}$. Once we set the chosen modular weights, their explicit forms read
\begin{subequations}
\label{eq:SeesawModelMassMtrices}
\begin{eqnarray} 
\label{eq:SeesawModelMassMtricesMe}
&&\!\!\!\!\!\!\!\!\!\!\!\!
  M_{e} = \begin{pmatrix}
      0 & \alpha^{\ell} Y^{(3)}_{\mathbf{\hat{1}'}} & -\beta^{\ell} Y^{(6)}_{\mathbf{2},2} \\
      -\alpha^{\ell} Y^{(3)}_{\mathbf{\hat{1}'}}  & 0 & \beta^{\ell} Y^{(6)}_{\mathbf{2},1}\\
      \gamma^{\ell} Y^{(2)}_{\mathbf{2},1} &\gamma^{\ell} Y^{(2)}_{\mathbf{2},2} &  0
      \end{pmatrix} v_{d}\,,\\
\label{eq:SeesawModelMassMtricesMnu}
&&\!\!\!\!\!\!\!\!\!\!\!\!
  M_{D} = \begin{pmatrix}
       -\alpha^\nu Y^{(2)}_{\mathbf{2},1} & \alpha^\nu Y^{(2)}_{\mathbf{2},2}  & \beta^\nu Y^{(5)}_{\mathbf{\hat{2}},1}  \\
       \alpha^\nu Y^{(2)}_{\mathbf{2},2} &  \alpha^\nu Y^{(2)}_{\mathbf{2},1} &\beta^\nu Y^{(5)}_{\mathbf{\hat{2}},2} \\
       \gamma^\nu Y^{(2)}_{\mathbf{2},1}  & \gamma^\nu Y^{(2)}_{\mathbf{2},2}  & 0 \\
       \end{pmatrix}v_{u}
       \,,~
  M_{N}= \begin{pmatrix}
       -\alpha^N Y^{(2)}_{\mathbf{2},1} & \alpha^N Y^{(2)}_{\mathbf{2},2}  & \beta^N Y^{(2)}_{\mathbf{2},1}  \\
       \alpha^N Y^{(2)}_{\mathbf{2},2} &  \alpha^N Y^{(2)}_{\mathbf{2},1} &\beta^N Y^{(2)}_{\mathbf{2},2}  \\
       \beta^N Y^{(2)}_{\mathbf{2},1}  & \beta^N Y^{(2)}_{\mathbf{2},2}  & 0 \\
       \end{pmatrix}\Lambda.
\end{eqnarray}
\end{subequations}
From the neutrino textures~\eqref{eq:SeesawModelMassMtricesMnu}, we see that the
light neutrino mass matrix $M_\nu$ is given by the standard seesaw formula,
\begin{equation}
   M_\nu ~=~ -M_D^T M_N^{-1} M_D \,.
\end{equation}
After minimizing the $\chi^2$ function~\eqref{eq:chi2}, we find that the best-fit point is determined by
\begin{eqnarray}
\nonumber
&&\braket{\tau}=0.197328+0.987258\I\,,\quad \tfrac{\beta^{\ell}}{\alpha^{\ell}}=447.984\,,\quad \tfrac{\gamma^{\ell}}{\alpha^{\ell}}=-32.5607\,,\quad \tfrac{\beta^\nu}{\alpha^\nu}=-3.63238\,,\\
&& \tfrac{\gamma^\nu}{\alpha^\nu}=-50.0039\,,~ \tfrac{\beta^N}{\alpha^N}=-26.6449\,, ~ \alpha^{\ell} v_{d}=1.96115~\text{MeV}\,,~\tfrac{(\alpha^\nu v_{u})^2}{\alpha^N\Lambda}=25.8397~\text{meV}\,.
\label{eq:Seesaw9parabf}
\end{eqnarray}
which lead to the lepton-sector predictions
\begin{eqnarray}
\nonumber&& \sin^{2}\theta^{\ell}_{12}=0.302991\,,\quad \sin^{2}\theta^{\ell}_{13}=0.02203\,,\quad \sin^{2}\theta^{\ell}_{23}=0.572022\,,\quad \delta_{\CP}^{\ell}=1.27887\pi\,,\\
\nonumber&& \alpha_{21}=1.27782\pi\,,\quad \alpha_{31}=0.418302\pi\,,\quad m_e/m_{\mu}=0.00473698\,,\quad m_{\mu}/m_{\tau}=0.0585681\,,\\
&&\frac{\Delta m_{21}^{2}}{\Delta m_{31}^{2}} =0.0295108\,,\quad  m_1=39.09~\text{meV}\,,\quad m_2=40.02~\text{meV}\,,\quad m_3=63.55~\text{meV}\,,
\end{eqnarray}
with $\chi^{2}_{\rm min}\approx 10^{-6}$. Note that in this case neutrino masses are predicted to be in normal ordering and all the predictions are compatible with the experimental data within $1\sigma$. Further, one can show that the effective mass $m_{\beta\beta}=22.48~\text{meV}$ is compatible with the current bound~\cite{KamLAND-Zen:2022tow}. In addition, it is interesting that the significant \CP violation phase 
$\delta_{\CP}^{\ell}=1.27887\pi$ 
is predicted entirely from the small deviation of the modulus vacuum $\braket{\tau}$ from the \CP-conserving boundary, $|\braket{\tau}|=1.00679\gtrsim 1$. On a lower key, the sum of neutrino masses $\sum_{i}m_{i}=142.66~\text{meV}$ is slightly larger than the upper bound of the Planck Collaboration results~\cite{Planck:2018vyg}, $\sum_{i}m_{i}<120~\text{meV}$.

%%%%%%%%%%%%%%%%%%%%%%%%%%%%%%%%%%%%
\subsection{Majorana neutrino model for Weinberg operator}
\label{sec:Weinbergneutino}

In the case of neutrino masses generated by the Weinberg operator, we find that at least 7 real parameters are needed to explain the measured values of lepton observables. Among approximately 300 models of this type with 7 parameters, around 15\% are found to be compatible with the experimental data. We present a sample lepton model in which the modular weights and representations of the matter fields are
\begin{align}
\nonumber
&E^c_D\oplus E^c_3 \sim \mathbf{2}\oplus \mathbf{\hat{1}}\,,\quad L_D\oplus L_3 \sim \mathbf{\hat{2}}\oplus \mathbf{\hat{1}} \,,\quad H_d \sim \mathbf{1}\,,~~H_u \sim \mathbf{1}\,.\\
&k_{E^c_D}=9/2\,,\quad k_{E^c_3}=k_{L_D}=k_{L_3}=3/2\,, \quad k_{H_d}=k_{H_u}=0\,.
\label{eq:WeinbergDef}
\end{align}
The corresponding modular invariant superpotential includes
\begin{equation}
\begin{aligned}
\mathcal{W}_{e}  &~=~ \left[\alpha^{\ell} \left(E_{D}^{c} L_D Y^{(5)}_{\mathbf{\hat{2}}}\right)_{\mathbf{1}}+ \beta^{\ell} \left(E_{D}^{c} L_3 Y^{(5)}_{\mathbf{\hat{2}}}\right)_{\mathbf{1}} +\gamma^{\ell}\left(E_{3}^{c} L_D Y^{(2)}_{\mathbf{2}}\right)_{\mathbf{1}} \right] H_d\,,\\
\mathcal{W}_{\nu}&~=~ \frac{1}{\Lambda} \left[\alpha^\nu \left(L_{D} L_D Y^{(2)}_{\mathbf{2}}\right)_{\mathbf{1}}+ \beta^\nu \left(L_{D} L_3 Y^{(2)}_{\mathbf{2}}\right)_{\mathbf{1}} +\beta^\nu \left(L_{3} L_D Y^{(2)}_{\mathbf{2}}\right)_{\mathbf{1}}\right] H_{u}H_{u}\,.
\end{aligned}
\end{equation}
From the defining properties~\eqref{eq:WeinbergDef} of our model, we observe that we must consider the mass textures $\mathcal{D}_{26}$ from Table~\ref{tab:DiracModels} for charged leptons, and $\mathcal{N}_{7}$ from Table~\ref{tab:MajoranaModels} for the neutrino masses.
After setting the corresponding weight assignments, the mass matrices read
\begin{subequations}
\label{eq:WeinbergModelMassMtrices}
\begin{align} 
\label{eq:WeinbergModelMassMtricesMe}
M_{e}&~=~ \begin{pmatrix}
\alpha^{\ell} Y^{(5)}_{\mathbf{\hat{2}},2} & \alpha^{\ell} Y^{(5)}_{\mathbf{\hat{2}},1} & \beta^{\ell} Y^{(5)}_{\mathbf{\hat{2}},2} \\
\alpha^{\ell} Y^{(5)}_{\mathbf{\hat{2}},1} & -\alpha^{\ell} Y^{(5)}_{\mathbf{\hat{2}},2} & -\beta^{\ell} Y^{(5)}_{\mathbf{\hat{2}},1}\\
-\gamma^{\ell} Y^{(2)}_{\mathbf{2},2} &\gamma^{\ell} Y^{(2)}_{\mathbf{2},1} &  0
\end{pmatrix} v_{d}\,,\\
\label{eq:WeinbergModelMassMtricesMnu}
M_\nu &~=~ 
\begin{pmatrix}
\alpha^\nu Y^{(2)}_{\mathbf{2},2} & \alpha^\nu Y^{(2)}_{\mathbf{2},1}  & -\beta^\nu Y^{(2)}_{\mathbf{2},2}  \\
\alpha^\nu Y^{(2)}_{\mathbf{2},1} &  -\alpha^\nu Y^{(2)}_{\mathbf{2},2} &\beta^\nu Y^{(2)}_{\mathbf{2},1} \\
-\beta^\nu Y^{(2)}_{\mathbf{2},2}  & \beta^\nu Y^{(2)}_{\mathbf{2},1}  & 0 \\
\end{pmatrix}\frac{v_{u}^{2}}{\Lambda}
\,.
\end{align}
\end{subequations}
The best-fit values of the input parameters are given by
\begin{equation}
\label{eq:Weinberg7parabf}
\begin{aligned}
\braket{\tau}&=0.143202+ 0.981206\I\,,\ &\beta^{\ell}/\alpha^{\ell}&=-3782.55\,,\    &\gamma^{\ell}/\alpha^{\ell}&=47.0628\,,\\
 \alpha^\nu/\beta^\nu&=-0.69875\,,      &\alpha^{\ell} v_{d}&=1.33421~\text{MeV}\,,\ &(\beta^\nu v_{u})^2/\Lambda&=34.8028~\text{meV}\,.
\end{aligned}
\end{equation}
The predictions of the lepton masses and flavor mixing parameters at this best-fit point are
\begin{eqnarray}
\nonumber&& \sin^{2}\theta^{\ell}_{12}=0.302725\,,\quad \sin^{2}\theta^{\ell}_{13}=0.0220538\,,\quad \sin^{2}\theta^{\ell}_{23}=0.656858\,,\quad \delta_{\CP}^{\ell}=1.46771\pi\,,\\
\nonumber&& \alpha_{21}=1.93205\pi\,,\quad \alpha_{31}=0.951594\pi\,,\quad m_e/m_{\mu}=0.00473702\,,\quad m_{\mu}/m_{\tau}=0.0585681\,,\\
&&\frac{\Delta m_{21}^{2}}{\Delta m_{31}^{2}} =0.0295217\,,\quad  m_1=30.15~\text{meV}\,,\quad m_2=31.35~\text{meV}\,,\quad m_3=58.47~\text{meV}\,,
\end{eqnarray}
with $\chi^{2}_{\rm min}\approx17.14$. We observe that neutrino masses are normal ordered, and that all predictions are compatible with the experimental data at $3\sigma$ level, except for $\sin^{2}\theta^{\ell}_{23}$, which is slightly outside the $3\sigma$ range. The \CP-violation phase $\delta_{\CP}^{\ell}$ is found close to $1.5\pi$. Moreover, the sum of neutrino masses $\sum_{i}m_{i}=119.96~\text{meV}$ is close to the upper bound of the Planck bound, $\sum_{i}m_{i}<120~\text{meV}$. Finally, the effective mass $m_{\beta\beta}=30.97~\text{meV}$ agrees with the current bound~\cite{KamLAND-Zen:2022tow} and can be confronted with data by future large-scale $0\nu\beta\beta$-decay experiments~\cite{nEXO:2021ujk}.

%%%%%%%%%%%%%%%%%%%%%%%%%%%%%%%%%%%%
\subsection{A complete model of quarks and leptons}
\label{sec:unified}

After having built some promising separate models of quarks or leptons, we are ready now to study whether our top-down motivated framework can yield a complete model of quarks and leptons. To keep contact with top-down constructions, we focus on complete models where all matter fields have the same representation assignments. We find that we need at least 16 input parameters to provide all of the 22 flavor parameters in the SM. Despite being challenged by some slightly more predictive models~\cite{Ding:2023ydy}, our model reveals the potential and simplicity of realistic modular flavor models based on $2D_3$. The details of the defining modular properties of all matter fields are summarized in Table~\ref{tab:UnifiedModels}.
\begin{table}[t!]
\centering
\begin{tabular}{|c|c|c|c|c|c|c|c|}\hline
& $d^c$ & $u^c$ & $Q$ & $E^c$ &  $L$ & $H_u$ & $H_d$  \\ \hline
$\rho$  & $\mathbf{\hat{2}}\oplus\mathbf{\hat{1}'}$ & $\mathbf{\hat{2}}\oplus\mathbf{\hat{1}'}$ & $\mathbf{\hat{2}}\oplus\mathbf{\hat{1}'}$ & $\mathbf{\hat{2}}\oplus\mathbf{\hat{1}'}$ &  $\mathbf{\hat{2}}\oplus\mathbf{\hat{1}'}$ & $\mathbf{1}'$ & $\mathbf{1}'$  \\ \hline
$k$  & $1/2,5/2$ & $5/2,1/2$ & $5/2,5/2$ & $-1/2,7/2$ &  $7/2,3/2$ & $0$ & $0$  \\ \hline
\end{tabular}
\caption{\label{tab:UnifiedModels} Transformation properties of quark and lepton fields under the binary dihedral group $2D_3$, and the modular weight assignments for a complete model of quarks and leptons, where neutrinos acquire masses via the Weinberg operator.}
\end{table}

By choice, all matter fields except for the Higgs fields have the same representation $\mathbf{\hat{2}}\oplus\mathbf{\hat{1}'}$ under $2D_3$. These assignments mean that the model can be naturally embedded in a Grand Unified Theory, and could in principle naturally arise from string theory compactified on a $\mathbbm T^2/\Z4$ orbifold~\cite{Baur:2023}.
Since the right-handed neutrino fields $N^c$ are not introduced in this model, neutrino masses originate from the Weinberg operator. In our complete model, due to the identical modular charges for all kinds of matter, the up-quark and down-quark mass matrices as well as the charged-lepton mass matrix are given by the texture $\mathcal{D}_6$ from Table~\ref{tab:DiracModels}. The neutrino mass matrix is given by the texture $\mathcal{N}_8$ from Table~\ref{tab:MajoranaModels}. In detail, our model yields
\begin{subequations}
\label{eq:UnifiedModelMassMtrices}
\begin{align} 
\label{eq:UnifiedModelMassMtricesMd}
M_{d}&~=~ \begin{pmatrix}
-\alpha^d Y^{(2)}_{\mathbf{2},1} & \alpha^d Y^{(2)}_{\mathbf{2},2}  & -\beta^d Y^{(2)}_{\mathbf{2},2}\\
\alpha^d Y^{(2)}_{\mathbf{2},2}  &  \alpha^d Y^{(2)}_{\mathbf{2},1} & \beta^d Y^{(2)}_{\mathbf{2},1} \\
-\gamma^d Y^{(4)}_{\mathbf{2},2} & \gamma^d Y^{(4)}_{\mathbf{2},1} &  \delta^d Y^{(4)}_{\mathbf{1}}
\end{pmatrix} v_{d}\,,\\
\label{eq:UnifiedModelMassMtricesMu}
M_{u}&~=~ \begin{pmatrix}
-\alpha^u_1 Y^{(4)}_{\mathbf{2},1}+\alpha^u_2 Y^{(4)}_{\mathbf{1}} & \alpha_1^u Y^{(4)}_{\mathbf{2},2}  & -\beta^u Y^{(4)}_{\mathbf{2},2}\\
\alpha_1^u Y^{(4)}_{\mathbf{2},2}  &  \alpha^u_1 Y^{(4)}_{\mathbf{2},1}+\alpha^u_2 Y^{(4)}_{\mathbf{1}} & \beta^u Y^{(4)}_{\mathbf{2},1} \\
-\gamma^u Y^{(2)}_{\mathbf{2},2} & \gamma^u Y^{(2)}_{\mathbf{2},1} &  0
\end{pmatrix} v_{u}\,,\\
\label{eq:UnifiedModelMassMtricesMe}
M_{e}&~=~ \begin{pmatrix}
-\alpha^{\ell} Y^{(2)}_{\mathbf{2},1} & \alpha^{\ell} Y^{(2)}_{\mathbf{2},2}  & 0\\
\alpha^{\ell} Y^{(2)}_{\mathbf{2},2}  &  \alpha^{\ell} Y^{(2)}_{\mathbf{2},1} & 0 \\
-\gamma^{\ell} Y^{(6)}_{\mathbf{2},2} &\gamma^{\ell} Y^{(6)}_{\mathbf{2},1} &  \delta^{\ell} Y^{(4)}_{\mathbf{1}}
\end{pmatrix} v_{d}\,,\\
\label{eq:UnifiedModelMassMtricesMnu}
M_{\nu}&~=~ \begin{pmatrix}
\alpha^\nu_1 Y^{(6)}_{\mathbf{2},2}+\alpha^\nu_2 Y^{(6)}_{\mathbf{1}'} & \alpha^\nu_1 Y^{(6)}_{\mathbf{2},1} & \beta^\nu Y^{(4)}_{\mathbf{2},1}  \\
\alpha^\nu_1 Y^{(6)}_{\mathbf{2},1} &  -\alpha^\nu_1 Y^{(6)}_{\mathbf{2},2}+\alpha^\nu_2 Y^{(6)}_{\mathbf{1}'} & \beta^\nu Y^{(4)}_{\mathbf{2},2}   \\
\beta^\nu Y^{(4)}_{\mathbf{2},1}  & \beta^\nu Y^{(4)}_{\mathbf{2},2}   & 0 \\
\end{pmatrix}\frac{v_u^2}{\Lambda}\,.
\end{align}
\end{subequations}

\begin{table}[h!]
\centering
\resizebox{0.5\textwidth}{!}{\small
\begin{tabular}{|c|c|c|c|}\hline \hline
\multicolumn{4}{|c|}{Input parameters} \\ \hline
$\re \langle\tau\rangle$ &$-0.0613689$ &$\im \langle\tau\rangle$ & $2.68637$ \\ \hline

$\alpha^u_1/\gamma^u$ & $-6422.49$ & $\gamma^{\ell}/\alpha^{\ell}$ & $17.0445$ \\ \hline
$\alpha^u_2/\gamma^u$ & $-6413.76$ & $\delta^{\ell}/\alpha^{\ell}$ & $-0.0808854$ \\ \hline
$\beta^u/\gamma^u$ & $4383.68$ & $\alpha^\nu_1/\beta^\nu$ & $2.3088$ \\ \hline
$\beta^d/\alpha^d$ & $20.3565$ & $\alpha^\nu_2/\beta^\nu$ & $-11784.2$ \\ \hline
$\gamma^d/\alpha^d$ & $-1040.08$ & $\alpha^{\ell} v_{d}[\text{MeV}]$ & $76.207$ \\ \hline 
$\delta^d/\alpha^d$ & $309.699$ & $\alpha^d v_{d}[\text{MeV}]$ & $0.891348$ \\ \hline	
$\gamma^u v_{u}[\text{MeV}]$ & $6.44174$ & $\beta^\nu v_u^2/\Lambda[\text{eV}]$ & $0.010066$ \\ \hline \hline 
\multicolumn{4}{c}{}\\
\multicolumn{4}{c}{}\\
\multicolumn{4}{c}{}\\
\multicolumn{4}{c}{}\\
\multicolumn{4}{c}{}\\[4mm]
\end{tabular}
} \hfill 
\resizebox{0.48\textwidth}{!}{\small
\begin{tabular}{|c|c|c|c|}\hline \hline
\multicolumn{4}{|c|}{Predicted flavor parameters} \\ \hline
\multicolumn{2}{|c|}{Quark sector} & \multicolumn{2}{c|}{Lepton sector} \\ \hline
$m_{u}/m_{c}$ & $0.00192463$ & $m_e/m_{\mu}$ & $0.00473731$ \\ \hline
$m_{c}/m_{t}$ & $0.00282265$ & $m_{\mu}/m_{\tau}$ & $0.058568$ \\ \hline
$m_{d}/m_{s}$ & $0.0505174$ & $\alpha_{21}/\pi$ & $0.901365$ \\ \hline
$m_{s}/m_{b}$ & $0.0182406$ & $\alpha_{31}/\pi$ & $0.526527$ \\ \hline
$\theta^{q}_{12}$ & $0.227464$ &
$\sin^{2}\theta^{\ell}_{12}$ & $0.31818$ \\ \hline
$\theta^{q}_{13}$ & $0.00339533$ & $\sin^{2}\theta^{\ell}_{13}$ & $0.021746$ \\ \hline
$\theta^{q}_{23}$ & $0.0403661$ & $\sin^{2}\theta^{\ell}_{23}$ & $0.527212$ \\ \hline
$\delta^{q}_{\CP}/{}^\circ$ & $69.1464$ & $\delta_{\CP}^{\ell}/\pi$ & $1.59044$ \\ \hline
\multicolumn{2}{c|}{} &	$m_1[\text{meV}]$ & $5.23272$  \\ \cline{3-4}
\multicolumn{2}{c|}{} &	$m_2[\text{meV}]$ & $10.0738$ \\ \cline{3-4}
\multicolumn{2}{c|}{} &	$m_3[\text{meV}]$ & $49.6888$ \\ \cline{3-4}
\multicolumn{2}{c|}{} &	$\Delta m_{21}^{2}/\Delta m_{31}^{2}$ & $0.030349$ \\ 
\cline{3-4}
\noalign{\vskip 7\arrayrulewidth}
\cline{3-4}
\end{tabular}}
\caption{\label{tab:CompleteModel}Best-fit values of the input parameters together with the corresponding predictions for flavor observables delivered by our complete model of quarks and leptons, defined in Table~\ref{tab:UnifiedModels}, endowed with a $2D_3$ modular flavor symmetry. The accuracy of this model is $\chi^{2}_\mathrm{min}\approx8.4$.}
\end{table}

In Table~\ref{tab:CompleteModel} we summarize the resulting best-fit input parameters, product of a $\chi^2$ minimization, along with the predictions for all flavor observables. The identified values yield an accuracy of $\chi^{2}_\mathrm{min}\approx8.4$, implying that all observables are found well within the experimentally allowed $3\sigma$ ranges. Besides, our model predicts, as in previous partial models, that
the leptonic Dirac \CP-violation phase $\delta_{\CP}^{\ell}$ is located around $1.5\pi$, and the sum of neutrino masses $\sum_{i}m_{i}=65~\text{meV}$ is somewhat lower than the current experimental upper bound.
Another interesting finding is that the best-fit of $\braket{\tau}$ is slightly off the \CP conserving boundary $\mathrm{Re}\tau=0$ and somewhat close to the critical point at $\I\infty$ in moduli space, similar to what was found previously in stringy models~\cite[Table 6]{Baur:2022hma}.
This small deviation in moduli space results in both a suitable \CP-violation phase for the quark sector and a larger \CP-violation phase for the lepton sector, which turn out to add up very close to $2\pi$, i.e. we find the intriguing approximate quark-lepton complementarity relation
\begin{equation}
  \delta_{\CP}^{q}+\delta_{\CP}^{\ell}~\approx~ 2\pi\,.
\end{equation}

%% file: Conclusion.tex
One interesting and unusual generalized finite modular group of small order that appears in some string compactifications is the binary dihedral group $2D_3\cong[12,1]$. Hence, unlike some other symmetries explored from the bottom-up perspective, there is additional top-down motivation to analyze the phenomenology arising from this modular group. In this work, we have systematically studied, from a bottom-up perspective, all the elements that are required to arrive at phenomenologically viable scenarios, in order to provide with the tools that top-down constructions shall eventually need.

We have first presented detailed group theoretical information for $2D_3$, whose representations suggest in particular a $2+1$-family structure. By applying the theory of VVMFs, we have then explicitly constructed all the modular-form multiplets of $2D_3$ with all admissible weights $k_Y\leq7$. They coincide with the VVMFs building doublets and singlets of the finite modular group $\Gamma'_4\cong S'_4$, as $2D_3\cong S_4'/(\Z2\x\Z2)$.

In our framework, we have adopted a number of top-down motivated assumptions, which differ slightly from typical bottom-up priors:
\begin{enumerate}[label=(\roman*)]
  \item Framework based on SUGRA, implying that $\mathcal{W}$ has a modular weight $-1$;
  \item Matter fields with fractional weights whenever possible;
  \item Higgs fields $H_{u/d}$ are allowed to be non-trivial $2D_3$ singlets; and
  \item Families are accommodated in $2+1$ structures instead of triplets.
\end{enumerate}
Beyond bringing many unexplored phenomenological possibilities, this serves as a bridge between the bottom-up and top-down approaches.

As a first general step towards phenomenology, we have comprehensively classified all Dirac and Majorana fermion mass textures that are consistent with a $2D_3$ finite modular theory of flavor. In our classification, aiming at minimalism, we have considered that neutrino masses can arise purely from Dirac terms, a type-I seesaw mechanism or the Weinberg operator. We identify all inequivalent mass textures with only the real numeric coefficients and modular weights as varying parameters in Tables~\ref{tab:DiracModels} and~\ref{tab:MajoranaModels}. This allows the interested readers to perform their own scans using the modular weights that are more convenient in their formalism.

Applying our classification, in section~\ref{sec:BenchmarkModels} we presented some results from a random scan among a large set of admissible combinations of fractional modular weights for matter fields. This scan allowed us first to find separate models with a reduced number of parameters for quarks and for leptons. Of course, only a complete model of quarks and leptons can faithfully be compared with observations. By scanning thousands of models, we identified a complete sample model, based on the $2D_3$ modular flavor symmetry, with only 16 parameters, that successfully provides all 22 flavor observables with $\chi^2\approx8.4$, cf.~Table~\ref{tab:CompleteModel}. Interestingly, neutrino masses are generated by the Weinberg operator, hence avoiding the need of right-handed neutrinos. Further, the $2D_3$ representations of all matter fields can be restricted to be identical, as observed in promising string-derived scenarios. The predictions of our complete model are also interesting; for instance, neutrino masses are relatively small with a total mass of only 65\,meV, and the \CP phases approximately satisfy a quark-lepton complementarity relation, $\delta_{\CP}^{q}+\delta_{\CP}^{\ell}\approx 2\pi$. The sample models of section~\ref{sec:BenchmarkModels} should be regarded as the main phenomenological result of our work.

The top-down motivated research in this paper opens up new options for both bottom-up and top-down flavor model building. First, as usual, it remains the challenge of finding mechanisms to stabilize the modulus at the best-fit point. Further, it would be interesting to investigate 
whether some of our models can be consistently embedded into some top-down framework, such as the heterotic string compactified on a $\mathbbm{T}^2/\Z4$ orbifold~\cite{Baur:2023}. This suggests the possibility of mixing the modular flavor symmetry $2D_3$ with a traditional flavor group, in an eclectic scheme, allowing us to fix the (canonical) structure of the K\"ahler potential and possibly induce further physical constraints. These tasks are left for future works.

%% file: app_hypergeometric.tex
The MLDEs in Eq.~\eqref{eq:MLDE} for 2-d and 3-d VVMFs can always be transformed into the generalized hypergeometric equation of the form~\cite{franc2016hypergeometric}
\begin{equation}
\label{eq:generalHyperDE}
\left[(\theta_K+\beta_1-1)\cdots (\theta_K+\beta_n-1) - K(\theta_K+\alpha_1)\cdots(\theta_K+\alpha_n)\right]f_i~=~0\,,
\end{equation}
where
\begin{equation}
	\theta_K~:=~K\frac{\dd}{\dd K}\qquad \text{with}\qquad
	K(\tau)~=~1728/j(\tau)\,.
\end{equation}
Here $j(\tau)$ is the Klein $j$-invariant, defined as~\cite{cohen2017modular}
\begin{equation}
\label{eq:j-Delta}
	j(\tau)~:=~\frac{E^3_4(\tau)}{\Delta(\tau)}\qquad \text{with}\qquad
	\Delta(\tau)~:=~\frac{E_4^3(\tau)-E_6^2(\tau)}{1728}\,.
\end{equation}
When the numbers $\beta_{1},\dots,\beta_{n}$ are distinct mod $\Z{}$, Eq.~\eqref{eq:generalHyperDE} has $n$ independent solutions which are given by the generalized hypergeometric series
\begin{equation}
\label{eq:solHyperEq}
   f_i~=~K^{1-\beta_i}~ {}_nF_{n-1}(1+\alpha_1-\beta_i,\dots,1+\alpha_n-\beta_i;1+\beta_1-\beta_i,\check{\dots} ,1+\beta_n-\beta_i; K)\,,
\end{equation}
where $\check{}$ denotes the omission of $1+\beta_i-\beta_i=1$. The generalized hypergeometric series ${}_nF_{n-1}$ is defined by the formula
\begin{equation}
\label{eq:HyperSeries}
{}_nF_{n-1}(a_1,\dots,a_n;b_1,\dots, b_{n-1}; z)~:=~ 
     \sum_{m\geq 0}^{\infty} \dfrac{\prod_{j=1}^{n} a^{(m)}_j}{\prod_{k=1}^{n-1} b^{(m)}_k} \dfrac{z^m}{m!}\,,\qquad m,j,k\in \mathbbm{N}\,,
\end{equation}
where $a^{(m)}_j$ (and $b^{(m)}_k$) is the rising factorial or Pochhammer symbol, defined for $a^{(m)}_j$ by
\begin{equation}
  a^{(m)}_j ~:=~ 
 \begin{cases}
 1                               &\qquad m=0 \,, \\
{a_j}({a_j}+1)\cdots({a_j}+m-1) &\qquad m\geq 1\,.
\end{cases}
\end{equation}
An analogous definition holds for $b^{(m)}_k$. In these expressions, $a_j,b_k \in\mathbbm{C}$. 

For the 2-d case, the matrix $\rho(T)$ corresponding to the unitary irreducible representation of $T\in\mathrm{SL}(2,\Z{})$ has eigenvalues $\e^{2\pi\I r_1}$ and $\e^{2\pi\I r_2}$, such that
\begin{equation}
    0 \leq r_1, r_2 < 1 \, , \qquad 
    r_1 + r_2 \in \frac{1}{6}\Z{} \,,\qquad 
    r_1 - r_2 \notin \Z{},~ \Z{} \pm \frac{1}{6}.
\end{equation}
In this case, the parameters $\beta_{1,2}$ and $\alpha_{1,2}$ can be solved by the indicial equation of the MLDE, and they solely rely on $r_1$ and $r_2$ according to
\begin{equation}
\beta_1~=~\frac{r_2-r_1}{2}+\frac{11}{12}\,,\qquad 
\beta_2~=~\frac{r_1-r_2}{2}+\frac{11}{12}\,,\qquad
\alpha_1~=~0\,,\qquad 
\alpha_2~=~\frac{1}{3}\,.
\end{equation}
Accordingly, the 2-d VVMFs $Y^{(k_0)}$ of minimal weight $k_0$ can be expressed   through the hypergeometric series~\eqref{eq:solHyperEq} as $Y^{(k_0)}=C_0(\eta^{2k_0}f_1, ~C_1 \eta^{2k_0}f_2)^\mathrm{T}$, where the overall coefficient $C_0$ can be generally taken as $1$ and the determination of the relative coefficient $C_1$ relies on the representation matrix $\rho(S)$. Note that the minimal weight $k_0$ here is determined to be $k_0=6(r_1+r_2)-1$.

%% file: app_Dirac.tex
\begin{table}[b!]
\begin{center}
\resizebox{1\textwidth}{!}{
\begin{tabular}{|c|c|c|}\hline\hline
Block & Representation configuration & Explicit form \\ 
label & $(\psi^c_a,\psi_b,H_{u/d})$  & of block mass texture\\ \hline 
$M_{DD_{1}}$ &$(\qa,\qa,\pa)$, $(\qb,\qb,\pb)$, $(\qa,\qb,\pd)$&
$\left(\begin{array}{cc} 
	\MDDi
\end{array}\right)$\\ \hline
$M_{{DD_{2}}^{\pm}}$ &$(\qa,\qb,\pa)^{+}$, $(\qa,\qa,\pc)^{+}$,$(\qb,\qb,\pd)^{-}$&
$\left(\begin{array}{cc} 
	\MDDiib
\end{array}\right)$\\ \hline
$M_{{DD_{3}}^{\pm}}$ &$(\qb,\qb,\pa)^{+}$,$(\qa,\qa,\pb)^{-}$, $(\qa,\qb,\pc)^{-}$&
$\left(\begin{array}{cc} 
	\MDDiiib
\end{array}\right)$\\ \hline
$M_{DD_{4}}$ &$(\qa,\qb,\pb)$, $(\qb,\qb,\pc)$, $(\qa,\qa,\pd)$&
$\left(\begin{array}{cc} 
	\MDDiv
\end{array}\right)$\\ \hline
$M_{D3_1}$ &$\begin{array}{c}(\qa,\pa,\pa),~(\qa,\pb,\pb),~(\qa,\pd,\pc),~(\qa,\pc,\pd),~\\
	(\qb,\pd,\pa),~(\qb,\pc,\pb),~(\qb,\pb,\pc),~(\qb,\pa,\pd)
\end{array}$ & $\left(\begin{array}{cc} \hspace{10pt}\MDTi \end{array}\right)$\\ \hline
$M_{{D3_{2}}^{\pm}}$ &$\begin{array}{c}(\qa,\pb,\pa)^{+},~(\qa,\pa,\pb)^{-},~(\qa,\pc,\pc)^{-},~(\qa,\pd,\pd)^{+},~\\
	(\qb,\pc,\pa)^{+},~(\qb,\pd,\pb)^{-},~(\qb,\pa,\pc)^{-},~(\qb,\pb,\pd)^{+}					
\end{array}$ & $\left(\begin{array}{cc} \hspace{10pt}\MDTiib \end{array}\right)$\\ \hline
$M_{{D3_{3}}^{\pm}}$ &$\begin{array}{c}(\qa,\pc,\pa)^{+},~(\qa,\pd,\pb)^{-},~(\qa,\pa,\pc)^{+},~(\qa,\pb,\pd)^{-},~\\
	(\qb,\pa,\pa)^{+},~(\qb,\pb,\pb)^{-},~(\qb,\pd,\pc)^{+},~(\qb,\pc,\pd)^{-}
\end{array}$ & $\left(\begin{array}{cc} \hspace{10pt}\MDTiiib \end{array}\right)$\\ \hline
$M_{D3_4}$ &$\begin{array}{c}(\qa,\pd,\pa),~(\qa,\pc,\pb),~(\qa,\pb,\pc),~(\qa,\pa,\pd),~\\
	(\qb,\pb,\pa),~(\qb,\pa,\pb),~(\qb,\pc,\pc),~(\qb,\pd,\pd)
\end{array}$ & $\left(\begin{array}{cc} \hspace{10pt}\MDTiv \end{array}\right)$\\ \hline
$M_{3D_1}$ &$\begin{array}{c}(\pa,\qa,\pa),~(\pb,\qa,\pb),~(\pd,\qa,\pc),~(\pc,\qa,\pd),~\\
	(\pd,\qb,\pa),~(\pc,\qb,\pb),~(\pb,\qb,\pc),~(\pa,\qb,\pd)
\end{array}$ & $\left(\begin{array}{cc} \MTDibis \end{array}\right)$\\ \hline
$M_{{3D_{2}}^{\pm}}$ &$\begin{array}{c}(\pb,\qa,\pa)^{+},~(\pa,\qa,\pb)^{-},~(\pc,\qa,\pc)^{-},~(\pd,\qa,\pd)^{+},~\\
	(\pc,\qb,\pa)^{+},~(\pd,\qb,\pb)^{-},~(\pa,\qb,\pc)^{-},~(\pb,\qb,\pd)^{+}
\end{array}$ & $\left(\begin{array}{cc} \MTDiibbis \end{array}\right)$\\ \hline
$M_{{3D_{3}}^{\pm}}$ &$\begin{array}{c}(\pc,\qa,\pa)^{+},~(\pd,\qa,\pb)^{-},~(\pa,\qa,\pc)^{+},~(\pb,\qa,\pd)^{-},~\\
	(\pa,\qb,\pa)^{+},~(\pb,\qb,\pb)^{-},~(\pd,\qb,\pc)^{+},~(\pc,\qb,\pd)^{-}
\end{array}$ & $\left(\begin{array}{cc} \MTDiiibbis \end{array}\right)$\\ \hline
$M_{3D_4}$ &$\begin{array}{c}(\pd,\qa,\pa),~(\pc,\qa,\pb),~(\pb,\qa,\pc),~(\pa,\qa,\pd),~\\
	(\pb,\qb,\pa),~(\pa,\qb,\pb),~(\pc,\qb,\pc),~(\pd,\qb,\pd)
\end{array}$ & $\left(\begin{array}{cc} \MTDivbis \end{array}\right)$\\
\hline\hline
\end{tabular}}
\caption{Admissible blocks in the Dirac mass matrix~\eqref{eq:MassMatrixInBlocks} for all possible $2D_3$ representation configurations of the involved fields $(\psi^c_a,\psi_b,H_{u/d})$, where $a,b$ are either $D$ or $3$. 
The superscript $+$ and $-$ in the representation configurations must be read as follows: 
the superscript $+$ in a representation configuration means that one must place $+$ ($-$) in all terms where the explicit form of the block displays $\pm$ ($\mp$). Analogously, $-$ in the representation configuration means a $-$ ($+$) when $\pm$ ($\mp$) is found. 
\label{tab:blocks_Dirac}}
\end{center}
\end{table}

We can compute the explicit form of the terms in Eq.~\eqref{eq:DiracSuperpotential} for the different field assignments by taking into account the $2D_3$ product rules, Eqs.~\eqref{eq:SingletProductRules}--\eqref{eq:DoubletProductRules}. All admissible independent mass blocks $M_{DD}$, $M_{D3}$ and $M_{3D}$ in Eq.~\eqref{eq:MassMatrixInBlocks} arising from considering the combinatorics of all $2D_3$ representations for every field, including the four different singlets for the Higgs fields, are shown in Table~\ref{tab:blocks_Dirac}. In this table we present the explicit form of the three independent blocks in Eq.~\eqref{eq:MassMatrixInBlocks}. In particular, as mentioned in section~\ref{sec:model}, the block $M_{33}$ is fixed once the three representation configurations $(\psi^c_a,\psi_b,H_{u/d})$ with $(a,b)=(D,D),(D,3)$ and $(3,D)$ and, hence, the respective blocks are chosen. For example, consider the configurations $(\rep2,\rep{\hat{2}},\rep{1'})$, $(\rep2,\rep{1'},\rep{1'})$ and $(\rep1,\rep{\hat{2}},\rep{1'})$. It follows that the configuration corresponding to $M_{33}$ must be $(\rep1,\rep{1'},\rep{1'})$, which would yield an $M_{33}$ being a modular form transforming in the $\rep{1'}$ representation of $2D_3$. \enlargethispage{\baselineskip}
All possible values of $M_{33}$ are $\delta Y^{(k_{33})}_{\pa}, \delta Y^{(k_{33})}_{\pb}, \delta Y^{(k_{33})}_{\pc}$ and $\delta Y^{(k_{33})}_{\pd}$, with $\delta\in\mathbbm{R}$.

In the second column of Table~\ref{tab:blocks_Dirac}, we gather together all representation configurations that lead to the same block structure up to a permutation of $\alpha_2$ and $\alpha_3$. Since $\alpha_i$ do not refer to any physical property, but simply to the order in which one considers the various terms arising in the products of the field representations, the blocks with permuted constants are equivalent. There are cases in the third column of our table where the sign of $\alpha_1$ or $\beta$ is relevant. The cases where $\pm$ or $\mp$ appear yield in principle two different blocks. However, this sign difference can be consistently redefined as long as both positive and negative coupling constants are considered, as we do in our scans. This eliminates in practice the duplicity of the blocks.

The weights $k_Y\in\{k_{DD},k_{3D},k_{D3},k_{33}\}$ of the modular forms in Table~\ref{tab:blocks_Dirac} are defined in section~\ref{sec:DiracYukawa}, below Eq.~\eqref{eq:DiracSuperpotential}. As they depend on the modular weights of the matter fields and these are fractional, it is clear that some of the resulting weights and required representations for modular forms are incompatible with the existent VVMFs of $2D_3$, Eqs.~\eqref{eq:MF-multiplets-2D3}. If no corresponding modular form multiplet exists at the given $k_Y$, then $Y^{(k_Y)}_{\rep{r}_Y}=0$. For instance, if $k_{\psi^c_D}+k_{\psi_D}=4$, then $k_{DD}:=k_{\psi^c_D}+k_{\psi_D}-1=3$ and thus $Y^{(k_{DD})}_{\rep2}=Y^{(k_{DD})}_{\rep{\hat{2}}}=0$.

Without setting explicitly the modular weights of matter fields, we can still build the full $3\x3$ Dirac mass textures $M_\psi$~\eqref{eq:MassMatrixInBlocks} by combining together the independent blocks from Table~\ref{tab:blocks_Dirac} and the corresponding $M_{33}$. We list all $4\x4\x4=64$ resulting textures in Table~\ref{tab:DiracModels} in terms of the full representation configuration $(\psi^c,\psi,H_{u/d})$. 

\begin{center}
\small \setlength\tabcolsep{4pt}
\begin{longtable}{|c|c|c|}
\hhline{===}
Model & Irrep configuration & \multirow{2}{*}{Dirac mass texture} \\ 
label & $(\psi^c,\psi,H_{u/d})$      &  \\
\hline\endfirsthead
\hline
Model & Irrep configuration & \multirow{2}{*}{Dirac mass texture} \\ 
label & $(\psi^c,\psi,H_{u/d})$      &  \\
\hline
\endhead
\multicolumn{3}{|r|}{continued...}\\
\hline
\endfoot
\endlastfoot
$\mathcal{D}_{1 }$&$\begin{array}{c}(\qa\oplus\pa,\qa\oplus\pa,\pa),\\(\qb\oplus\pc,\qb\oplus\pc,\pb),\\
	(\qa\oplus\pa,\qb\oplus\pc,\pd)	
\end{array}$ & $\left(\begin{array}{ccc}
	\MDDi & \MDTi \\
	\MTDi & \MTTi  \\
\end{array}\right)$ \\ \hline
$\mathcal{D}_{2 }$&$\begin{array}{c}(\qa\oplus\pb,\qa\oplus\pa,\pa),\\(\qb\oplus\pd,\qb\oplus\pc,\pb),\\
	(\qa\oplus\pb,\qb\oplus\pc,\pd)
\end{array}$ & $\left(\begin{array}{ccc}
	\MDDi & \MDTi \\
	\MTDii & \MTTii  \\
\end{array}\right)$ \\ \hline
$\mathcal{D}_{3 }$&$\begin{array}{c}(\qa\oplus\pc,\qa\oplus\pa,\pa),\\(\qb\oplus\pb,\qb\oplus\pc,\pb),\\
	(\qa\oplus\pc,\qb\oplus\pc,\pd)
\end{array}$ & $\left(\begin{array}{ccc}
	\MDDi & \MDTi \\
	\MTDiii & \MTTiv  \\
\end{array}\right)$ \\ \hline
$\mathcal{D}_{4 }$&$\begin{array}{c}(\qa\oplus\pd,\qa\oplus\pa,\pa),\\(\qb\oplus\pa,\qb\oplus\pc,\pb),\\
	(\qa\oplus\pd,\qb\oplus\pc,\pd)
\end{array}$ & $\left(\begin{array}{ccc}
	\MDDi & \MDTi \\
	\MTDiv & \MTTiii  \\
\end{array}\right)$ \\ \hline
$\mathcal{D}_{5 }$&$\begin{array}{c}(\qa\oplus\pa,\qa\oplus\pb,\pa),\\(\qb\oplus\pc,\qb\oplus\pd,\pb),\\
	(\qa\oplus\pa,\qb\oplus\pd,\pd)
\end{array}$ & $\left(\begin{array}{ccc}
	\MDDi & \MDTii \\
	\MTDi & \MTTii  \\
\end{array}\right)$ \\ \hline
$\mathcal{D}_{6 }$&$\begin{array}{c}(\qa\oplus\pb,\qa\oplus\pb,\pa),\\(\qb\oplus\pd,\qb\oplus\pd,\pb),\\
	(\qa\oplus\pb,\qb\oplus\pd,\pd)
\end{array}$ & $\left(\begin{array}{ccc}
	\MDDi & \MDTii \\
	\MTDii & \MTTi  \\
\end{array}\right)$ \\ \hline
$\mathcal{D}_{7 }$&$\begin{array}{c}(\qa\oplus\pc,\qa\oplus\pb,\pa),\\(\qb\oplus\pb,\qb\oplus\pd,\pb),\\
	(\qa\oplus\pc,\qb\oplus\pd,\pd)
\end{array}$ & $\left(\begin{array}{ccc}
	\MDDi & \MDTii \\
	\MTDiii & \MTTiii  \\
\end{array}\right)$ \\ \hline
$\mathcal{D}_{8 }$&$\begin{array}{c}(\qa\oplus\pd,\qa\oplus\pb,\pa),\\(\qb\oplus\pa,\qb\oplus\pd,\pb),\\
	(\qa\oplus\pd,\qb\oplus\pd,\pd)
\end{array}$ & $\left(\begin{array}{ccc}
	\MDDi & \MDTii \\
	\MTDiv & \MTTiv  \\
\end{array}\right)$ \\ \hline
$\mathcal{D}_{9 }$&$\begin{array}{c}(\qa\oplus\pa,\qa\oplus\pc,\pa),\\(\qb\oplus\pc,\qb\oplus\pb,\pb),\\
	(\qa\oplus\pa,\qb\oplus\pb,\pd)
\end{array}$ & $\left(\begin{array}{ccc}
	\MDDi & \MDTiii \\
	\MTDi & \MTTiv  \\
\end{array}\right)$ \\ \hline
$\mathcal{D}_{10}$&$\begin{array}{c}(\qa\oplus\pb,\qa\oplus\pc,\pa),\\(\qb\oplus\pd,\qb\oplus\pb,\pb),\\
	(\qa\oplus\pb,\qb\oplus\pb,\pd)
\end{array}$ & $\left(\begin{array}{ccc}
	\MDDi & \MDTiii \\
	\MTDii & \MTTiii  \\
\end{array}\right)$ \\ \hline
$\mathcal{D}_{11}$&$\begin{array}{c}(\qa\oplus\pc,\qa\oplus\pc,\pa),\\(\qb\oplus\pb,\qb\oplus\pb,\pb),\\
	(\qa\oplus\pc,\qb\oplus\pb,\pd)
\end{array}$ & $\left(\begin{array}{ccc}
	\MDDi & \MDTiii \\
	\MTDiii & \MTTii  \\
\end{array}\right)$ \\ \hline
$\mathcal{D}_{12}$&$\begin{array}{c}(\qa\oplus\pd,\qa\oplus\pc,\pa),\\(\qb\oplus\pa,\qb\oplus\pb,\pb),\\
	(\qa\oplus\pd,\qb\oplus\pb,\pd)
\end{array}$ & $\left(\begin{array}{ccc}
	\MDDi & \MDTiii \\
	\MTDiv & \MTTi  \\
\end{array}\right)$ \\ \hline
$\mathcal{D}_{13}$&$\begin{array}{c}(\qa\oplus\pa,\qa\oplus\pd,\pa),\\(\qb\oplus\pc,\qb\oplus\pa,\pb),\\
	(\qa\oplus\pa,\qb\oplus\pa,\pd)
\end{array}$ & $\left(\begin{array}{ccc}
	\MDDi & \MDTiv \\
	\MTDi & \MTTiii  \\
\end{array}\right)$ \\ \hline
$\mathcal{D}_{14}$&$\begin{array}{c}(\qa\oplus\pb,\qa\oplus\pd,\pa),\\(\qb\oplus\pd,\qb\oplus\pa,\pb),\\
	(\qa\oplus\pb,\qb\oplus\pa,\pd)
\end{array}$ & $\left(\begin{array}{ccc}
	\MDDi & \MDTiv \\
	\MTDii & \MTTiv  \\
\end{array}\right)$ \\ \hline
$\mathcal{D}_{15}$&$\begin{array}{c}(\qa\oplus\pc,\qa\oplus\pd,\pa),\\(\qb\oplus\pb,\qb\oplus\pa,\pb),\\
	(\qa\oplus\pc,\qb\oplus\pa,\pd)
\end{array}$ & $\left(\begin{array}{ccc}
	\MDDi & \MDTiv \\
	\MTDiii & \MTTi  \\
\end{array}\right)$ \\ \hline
$\mathcal{D}_{16}$&$\begin{array}{c}(\qa\oplus\pd,\qa\oplus\pd,\pa),\\(\qb\oplus\pa,\qb\oplus\pa,\pb),\\
	(\qa\oplus\pd,\qb\oplus\pa,\pd)
\end{array}$ & $\left(\begin{array}{ccc}
	\MDDi & \MDTiv \\
	\MTDiv & \MTTii  \\
\end{array}\right)$ \\ \hline
$\mathcal{D}_{17}$&$\begin{array}{c}(\qa\oplus\pd,\qb\oplus\pa,\pa),\\(\qa\oplus\pd,\qa\oplus\pd,\pc),\\
	(\qb\oplus\pa,\qb\oplus\pa,\pd)
\end{array}$ & $\left(\begin{array}{ccc}
	\MDDii & \MDTi \\
	\MTDi & \MTTiii  \\
\end{array}\right)$ \\ \hline
$\mathcal{D}_{18}$&$\begin{array}{c}(\qa\oplus\pc,\qb\oplus\pa,\pa),\\(\qa\oplus\pc,\qa\oplus\pd,\pc),\\
	(\qb\oplus\pb,\qb\oplus\pa,\pd)
\end{array}$ & $\left(\begin{array}{ccc}
	\MDDii & \MDTi \\
	\MTDii & \MTTiv  \\
\end{array}\right)$ \\ \hline
$\mathcal{D}_{19}$&$\begin{array}{c}(\qa\oplus\pa,\qb\oplus\pa,\pa),\\(\qa\oplus\pa,\qa\oplus\pd,\pc),\\
	(\qb\oplus\pc,\qb\oplus\pa,\pd)
\end{array}$ & $\left(\begin{array}{ccc}
	\MDDii & \MDTi \\
	\MTDiii & \MTTi  \\
\end{array}\right)$ \\ \hline
$\mathcal{D}_{20}$&$\begin{array}{c}(\qa\oplus\pb,\qb\oplus\pa,\pa),\\(\qa\oplus\pb,\qa\oplus\pd,\pc),\\
	(\qb\oplus\pd,\qb\oplus\pa,\pd)
\end{array}$ & $\left(\begin{array}{ccc}
	\MDDii & \MDTi \\
	\MTDiv & \MTTii  \\
\end{array}\right)$ \\ \hline
$\mathcal{D}_{21}$&$\begin{array}{c}(\qa\oplus\pd,\qb\oplus\pb,\pa),\\(\qa\oplus\pd,\qa\oplus\pc,\pc),\\
	(\qb\oplus\pa,\qb\oplus\pb,\pd)
\end{array}$ & $\left(\begin{array}{ccc}
	\MDDii & \MDTii \\
	\MTDi & \MTTiv  \\
\end{array}\right)$ \\ \hline
$\mathcal{D}_{22}$&$\begin{array}{c}(\qa\oplus\pc,\qb\oplus\pb,\pa),\\(\qa\oplus\pc,\qa\oplus\pc,\pc),\\
	(\qb\oplus\pb,\qb\oplus\pb,\pd)
\end{array}$ & $\left(\begin{array}{ccc}
	\MDDii & \MDTii \\
	\MTDii & \MTTiii  \\
\end{array}\right)$ \\ \hline
$\mathcal{D}_{23}$&$\begin{array}{c}(\qa\oplus\pa,\qb\oplus\pb,\pa),\\(\qa\oplus\pa,\qa\oplus\pc,\pc),\\
	(\qb\oplus\pc,\qb\oplus\pb,\pd)
\end{array}$ & $\left(\begin{array}{ccc}
	\MDDii & \MDTii \\
	\MTDiii & \MTTii  \\
\end{array}\right)$ \\ \hline
$\mathcal{D}_{24}$&$\begin{array}{c}(\qa\oplus\pb,\qb\oplus\pb,\pa),\\(\qa\oplus\pb,\qa\oplus\pc,\pc),\\
	(\qb\oplus\pd,\qb\oplus\pb,\pd)
\end{array}$ & $\left(\begin{array}{ccc}
	\MDDii & \MDTii \\
	\MTDiv & \MTTi  \\
\end{array}\right)$ \\ \hline
$\mathcal{D}_{25}$&$\begin{array}{c}(\qa\oplus\pd,\qb\oplus\pc,\pa),\\(\qa\oplus\pd,\qa\oplus\pa,\pc),\\
	(\qb\oplus\pa,\qb\oplus\pc,\pd)
\end{array}$ & $\left(\begin{array}{ccc}
	\MDDii & \MDTiii \\
	\MTDi & \MTTi  \\
\end{array}\right)$ \\ \hline
$\mathcal{D}_{26}$&$\begin{array}{c}(\qa\oplus\pc,\qb\oplus\pc,\pa),\\(\qa\oplus\pc,\qa\oplus\pa,\pc),\\
	(\qb\oplus\pb,\qb\oplus\pc,\pd)
\end{array}$ & $\left(\begin{array}{ccc}
	\MDDii & \MDTiii \\
	\MTDii & \MTTii  \\
\end{array}\right)$ \\ \hline
$\mathcal{D}_{27}$&$\begin{array}{c}(\qa\oplus\pa,\qb\oplus\pc,\pa),\\(\qa\oplus\pa,\qa\oplus\pa,\pc),\\
	(\qb\oplus\pc,\qb\oplus\pc,\pd)
\end{array}$ & $\left(\begin{array}{ccc}
	\MDDii & \MDTiii \\
	\MTDiii & \MTTiv  \\
\end{array}\right)$ \\ \hline
$\mathcal{D}_{28}$&$\begin{array}{c}(\qa\oplus\pb,\qb\oplus\pc,\pa),\\(\qa\oplus\pb,\qa\oplus\pa,\pc),\\
	(\qb\oplus\pd,\qb\oplus\pc,\pd)
\end{array}$ & $\left(\begin{array}{ccc}
	\MDDii & \MDTiii \\
	\MTDiv & \MTTiii  \\
\end{array}\right)$ \\ \hline
$\mathcal{D}_{29}$&$\begin{array}{c}(\qa\oplus\pd,\qb\oplus\pd,\pa),\\(\qa\oplus\pd,\qa\oplus\pb,\pc),\\
	(\qb\oplus\pa,\qb\oplus\pd,\pd)
\end{array}$ & $\left(\begin{array}{ccc}
	\MDDii & \MDTiv \\
	\MTDi & \MTTii  \\
\end{array}\right)$ \\ \hline
$\mathcal{D}_{30}$&$\begin{array}{c}(\qa\oplus\pc,\qb\oplus\pd,\pa),\\(\qa\oplus\pc,\qa\oplus\pb,\pc),\\
	(\qb\oplus\pb,\qb\oplus\pd,\pd)
\end{array}$ & $\left(\begin{array}{ccc}
	\MDDii & \MDTiv \\
	\MTDii & \MTTi  \\
\end{array}\right)$ \\ \hline
$\mathcal{D}_{31}$&$\begin{array}{c}(\qa\oplus\pa,\qb\oplus\pd,\pa),\\(\qa\oplus\pa,\qa\oplus\pb,\pc),\\
	(\qb\oplus\pc,\qb\oplus\pd,\pd)
\end{array}$ & $\left(\begin{array}{ccc}
	\MDDii & \MDTiv \\
	\MTDiii & \MTTiii  \\
\end{array}\right)$ \\ \hline
$\mathcal{D}_{32}$&$\begin{array}{c}(\qa\oplus\pb,\qb\oplus\pd,\pa),\\(\qb\oplus\pd,\qb\oplus\pd,\pd),\\
	(\qa\oplus\pb,\qa\oplus\pb,\pc)
\end{array}$ & $\left(\begin{array}{ccc}
	\MDDii & \MDTiv \\
	\MTDiv & \MTTiv  \\
\end{array}\right)$ \\ \hline		
$\mathcal{D}_{33}$&$\begin{array}{c}(\qb\oplus\pd,\qb\oplus\pd,\pa),\\(\qa\oplus\pb,\qa\oplus\pb,\pb),\\
	(\qa\oplus\pb,\qb\oplus\pd,\pc)
\end{array}$ & $\left(\begin{array}{ccc}
	\MDDiii & \MDTi \\
	\MTDi & \MTTii  \\
\end{array}\right)$ \\ \hline
$\mathcal{D}_{34}$&$\begin{array}{c}(\qb\oplus\pc,\qb\oplus\pd,\pa),\\(\qa\oplus\pa,\qa\oplus\pb,\pb),\\
	(\qa\oplus\pa,\qb\oplus\pd,\pc)
\end{array}$ & $\left(\begin{array}{ccc}
	\MDDiii & \MDTi \\
	\MTDii & \MTTi  \\
\end{array}\right)$ \\ \hline
$\mathcal{D}_{35}$&$\begin{array}{c}(\qb\oplus\pa,\qb\oplus\pd,\pa),\\(\qa\oplus\pd,\qb\oplus\pd,\pc),\\
	(\qa\oplus\pd,\qa\oplus\pb,\pb)
\end{array}$ & $\left(\begin{array}{ccc}
	\MDDiii & \MDTi \\
	\MTDiii & \MTTiii  \\
\end{array}\right)$ \\ \hline
$\mathcal{D}_{36}$&$\begin{array}{c}(\qb\oplus\pb,\qb\oplus\pd,\pa),\\(\qa\oplus\pc,\qb\oplus\pd,\pc),\\
	(\qa\oplus\pc,\qa\oplus\pb,\pb)
\end{array}$ & $\left(\begin{array}{ccc}
	\MDDiii & \MDTi \\
	\MTDiv & \MTTiv  \\
\end{array}\right)$ \\ \hline			
$\mathcal{D}_{37}$&$\begin{array}{c}(\qb\oplus\pd,\qb\oplus\pc,\pa),\\(\qa\oplus\pb,\qa\oplus\pa,\pb),\\
	(\qa\oplus\pb,\qb\oplus\pc,\pc)
\end{array}$ & $\left(\begin{array}{ccc}
	\MDDiii & \MDTii \\
	\MTDi & \MTTi  \\
\end{array}\right)$ \\ \hline
$\mathcal{D}_{38}$&$\begin{array}{c}(\qb\oplus\pc,\qb\oplus\pc,\pa),\\(\qa\oplus\pa,\qa\oplus\pa,\pb),\\
	(\qa\oplus\pa,\qb\oplus\pc,\pc)
\end{array}$ & $\left(\begin{array}{ccc}
	\MDDiii & \MDTii \\
	\MTDii & \MTTii  \\
\end{array}\right)$ \\ \hline
$\mathcal{D}_{39}$&$\begin{array}{c}(\qb\oplus\pa,\qb\oplus\pc,\pa),\\(\qa\oplus\pd,\qb\oplus\pc,\pc),\\
	(\qa\oplus\pd,\qa\oplus\pa,\pb)
\end{array}$ & $\left(\begin{array}{ccc}
	\MDDiii & \MDTii \\
	\MTDiii & \MTTiv  \\
\end{array}\right)$ \\ \hline
$\mathcal{D}_{40}$&$\begin{array}{c}(\qb\oplus\pb,\qb\oplus\pc,\pa),\\(\qa\oplus\pc,\qb\oplus\pc,\pc),\\
	(\qa\oplus\pc,\qa\oplus\pa,\pb)
\end{array}$ & $\left(\begin{array}{ccc}
	\MDDiii & \MDTii \\
	\MTDiv & \MTTiii  \\
\end{array}\right)$ \\ \hline	$\mathcal{D}_{41}$&$\begin{array}{c}(\qb\oplus\pd,\qb\oplus\pa,\pa),\\(\qa\oplus\pb,\qa\oplus\pd,\pb),\\
	(\qa\oplus\pb,\qb\oplus\pa,\pc)
\end{array}$ & $\left(\begin{array}{ccc}
	\MDDiii & \MDTiii \\
	\MTDi & \MTTiii  \\
\end{array}\right)$ \\ \hline
$\mathcal{D}_{42}$&$\begin{array}{c}(\qb\oplus\pc,\qb\oplus\pa,\pa),\\(\qa\oplus\pa,\qa\oplus\pd,\pb),\\
	(\qa\oplus\pa,\qb\oplus\pa,\pc)
\end{array}$ & $\left(\begin{array}{ccc}
	\MDDiii & \MDTiii \\
	\MTDii & \MTTiv  \\
\end{array}\right)$ \\ \hline
$\mathcal{D}_{43}$&$\begin{array}{c}(\qb\oplus\pa,\qb\oplus\pa,\pa),\\(\qa\oplus\pd,\qb\oplus\pa,\pc),\\
	(\qa\oplus\pd,\qa\oplus\pd,\pb)
\end{array}$ & $\left(\begin{array}{ccc}
	\MDDiii & \MDTiii \\
	\MTDiii & \MTTi  \\
\end{array}\right)$ \\ \hline
$\mathcal{D}_{44}$&$\begin{array}{c}(\qb\oplus\pb,\qb\oplus\pa,\pa),\\(\qa\oplus\pc,\qb\oplus\pa,\pc),\\
	(\qa\oplus\pc,\qa\oplus\pd,\pb)
\end{array}$ & $\left(\begin{array}{ccc}
	\MDDiii & \MDTiii \\
	\MTDiv & \MTTii  \\
\end{array}\right)$ \\ \hline
$\mathcal{D}_{45}$&$\begin{array}{c}(\qb\oplus\pd,\qb\oplus\pb,\pa),\\(\qa\oplus\pb,\qa\oplus\pc,\pb),\\
	(\qa\oplus\pb,\qb\oplus\pb,\pc)
\end{array}$ & $\left(\begin{array}{ccc}
	\MDDiii & \MDTiv \\
	\MTDi & \MTTiv  \\
\end{array}\right)$ \\ \hline
$\mathcal{D}_{46}$&$\begin{array}{c}(\qb\oplus\pc,\qb\oplus\pb,\pa),\\(\qa\oplus\pa,\qa\oplus\pc,\pb),\\
	(\qa\oplus\pa,\qb\oplus\pb,\pc)
\end{array}$ & $\left(\begin{array}{ccc}
	\MDDiii & \MDTiv \\
	\MTDii & \MTTiii  \\
\end{array}\right)$ \\ \hline
$\mathcal{D}_{47}$&$\begin{array}{c}(\qb\oplus\pa,\qb\oplus\pb,\pa),\\(\qa\oplus\pd,\qb\oplus\pb,\pc),\\
	(\qa\oplus\pd,\qa\oplus\pc,\pb)
\end{array}$ & $\left(\begin{array}{ccc}
	\MDDiii & \MDTiv \\
	\MTDiii & \MTTii  \\
\end{array}\right)$ \\ \hline
$\mathcal{D}_{48}$&$\begin{array}{c}(\qb\oplus\pb,\qb\oplus\pb,\pa),\\(\qa\oplus\pc,\qb\oplus\pb,\pc),\\
	(\qa\oplus\pc,\qa\oplus\pc,\pb)
\end{array}$ & $\left(\begin{array}{ccc}
	\MDDiii & \MDTiv \\
	\MTDiv & \MTTi  \\
\end{array}\right)$ \\ \hline
$\mathcal{D}_{49}$&$\begin{array}{c}(\qa\oplus\pc,\qb\oplus\pb,\pb),\\(\qb\oplus\pb,\qb\oplus\pb,\pc),\\
	(\qa\oplus\pc,\qa\oplus\pc,\pd)
\end{array}$ & $\left(\begin{array}{ccc}
	\MDDiv & \MDTi \\
	\MTDi & \MTTiv  \\
\end{array}\right)$ \\ \hline
$\mathcal{D}_{50}$&$\begin{array}{c}(\qa\oplus\pd,\qb\oplus\pb,\pb),\\(\qb\oplus\pa,\qb\oplus\pb,\pc),\\
	(\qa\oplus\pd,\qa\oplus\pc,\pd)
\end{array}$ & $\left(\begin{array}{ccc}
	\MDDiv & \MDTi \\
	\MTDii & \MTTiii  \\
\end{array}\right)$ \\ \hline
$\mathcal{D}_{51 }$&$\begin{array}{c}(\qa\oplus\pb,\qb\oplus\pb,\pb),\\(\qb\oplus\pd,\qb\oplus\pb,\pc),\\
	(\qa\oplus\pb,\qa\oplus\pc,\pd)
\end{array}$ & $\left(\begin{array}{ccc}
	\MDDiv & \MDTi \\
	\MTDiii & \MTTii  \\
\end{array}\right)$ \\ \hline			
$\mathcal{D}_{52}$&$\begin{array}{c}(\qa\oplus\pa,\qb\oplus\pb,\pb),\\(\qa\oplus\pa,\qa\oplus\pc,\pd),\\
	(\qb\oplus\pc,\qb\oplus\pb,\pc)
\end{array}$ & $\left(\begin{array}{ccc}
	\MDDiv & \MDTi \\
	\MTDiv & \MTTi  \\
\end{array}\right)$ \\ \hline
$\mathcal{D}_{53}$&$\begin{array}{c}(\qa\oplus\pc,\qb\oplus\pa,\pb),\\(\qb\oplus\pb,\qb\oplus\pa,\pc),\\
	(\qa\oplus\pc,\qa\oplus\pd,\pd)
\end{array}$ & $\left(\begin{array}{ccc}
	\MDDiv & \MDTii \\
	\MTDi & \MTTiii  \\
\end{array}\right)$ \\ \hline
$\mathcal{D}_{54}$&$\begin{array}{c}(\qa\oplus\pd,\qb\oplus\pa,\pb),\\(\qb\oplus\pa,\qb\oplus\pa,\pc),\\
	(\qa\oplus\pd,\qa\oplus\pd,\pd)
\end{array}$ & $\left(\begin{array}{ccc}
	\MDDiv & \MDTii \\
	\MTDii & \MTTiv  \\
\end{array}\right)$ \\ \hline
$\mathcal{D}_{55}$&$\begin{array}{c}(\qa\oplus\pb,\qb\oplus\pa,\pb),\\(\qb\oplus\pd,\qb\oplus\pa,\pc),\\
	(\qa\oplus\pb,\qa\oplus\pd,\pd)
\end{array}$ & $\left(\begin{array}{ccc}
	\MDDiv & \MDTii \\
	\MTDiii & \MTTi  \\
\end{array}\right)$ \\ \hline
$\mathcal{D}_{56}$&$\begin{array}{c}(\qa\oplus\pa,\qb\oplus\pa,\pb),\\(\qb\oplus\pc,\qb\oplus\pa,\pc),\\
	(\qa\oplus\pa,\qa\oplus\pd,\pd)
\end{array}$ & $\left(\begin{array}{ccc}
	\MDDiv & \MDTii \\
	\MTDiv & \MTTii  \\
\end{array}\right)$ \\ \hline
$\mathcal{D}_{57}$&$\begin{array}{c}(\qa\oplus\pc,\qb\oplus\pd,\pb),\\(\qb\oplus\pb,\qb\oplus\pd,\pc),\\
	(\qa\oplus\pc,\qa\oplus\pb,\pd)
\end{array}$ & $\left(\begin{array}{ccc}
	\MDDiv & \MDTiii \\
	\MTDi & \MTTii  \\
\end{array}\right)$ \\ \hline
$\mathcal{D}_{58}$&$\begin{array}{c}(\qa\oplus\pd,\qb\oplus\pd,\pb),\\(\qb\oplus\pa,\qb\oplus\pd,\pc),\\
	(\qa\oplus\pd,\qa\oplus\pb,\pd)
\end{array}$ & $\left(\begin{array}{ccc}
	\MDDiv & \MDTiii \\
	\MTDii & \MTTi  \\
\end{array}\right)$ \\ \hline
$\mathcal{D}_{59}$&$\begin{array}{c}(\qa\oplus\pb,\qb\oplus\pd,\pb),\\(\qb\oplus\pd,\qb\oplus\pd,\pc),\\
	(\qa\oplus\pb,\qa\oplus\pb,\pd)
\end{array}$ & $\left(\begin{array}{ccc}
	\MDDiv & \MDTiii \\
	\MTDiii & \MTTiii  \\
\end{array}\right)$ \\ \hline
$\mathcal{D}_{60}$&$\begin{array}{c}(\qa\oplus\pa,\qb\oplus\pd,\pb),\\(\qa\oplus\pa,\qa\oplus\pb,\pd),\\
	(\qb\oplus\pc,\qb\oplus\pd,\pc)
\end{array}$ & $\left(\begin{array}{ccc}
	\MDDiv & \MDTiii \\
	\MTDiv & \MTTiv  \\
\end{array}\right)$ \\ \hline
$\mathcal{D}_{61}$&$\begin{array}{c}(\qa\oplus\pc,\qb\oplus\pc,\pb),\\(\qb\oplus\pb,\qb\oplus\pc,\pc),\\
	(\qa\oplus\pc,\qa\oplus\pa,\pd)
\end{array}$ & $\left(\begin{array}{ccc}
	\MDDiv & \MDTiv \\
	\MTDi & \MTTi  \\
\end{array}\right)$ \\ \hline
$\mathcal{D}_{62}$&$\begin{array}{c}(\qa\oplus\pd,\qb\oplus\pc,\pb),\\(\qb\oplus\pa,\qb\oplus\pc,\pc),\\
	(\qa\oplus\pd,\qa\oplus\pa,\pd)
\end{array}$ & $\left(\begin{array}{ccc}
	\MDDiv & \MDTiv \\
	\MTDii & \MTTii  \\
\end{array}\right)$ \\ \hline
$\mathcal{D}_{63}$&$\begin{array}{c}(\qa\oplus\pb,\qb\oplus\pc,\pb),\\(\qb\oplus\pd,\qb\oplus\pc,\pc),\\
	(\qa\oplus\pb,\qa\oplus\pa,\pd)
\end{array}$ & $\left(\begin{array}{ccc}
	\MDDiv & \MDTiv \\
	\MTDiii & \MTTiv  \\
\end{array}\right)$ \\ \hline				
$\mathcal{D}_{64}$&$\begin{array}{c}(\qa\oplus\pa,\qb\oplus\pc,\pb),\\(\qa\oplus\pa,\qa\oplus\pa,\pd),\\
	(\qb\oplus\pc,\qb\oplus\pc,\pc)
\end{array}$ & $\left(\begin{array}{ccc}
	\MDDiv & \MDTiv \\
	\MTDiv & \MTTiii  \\
\end{array}\right)$ \\ \hhline{===} 
\caption{\label{tab:DiracModels} All admissible Dirac mass textures
in models based on a $2D_3$ modular flavor symmetry, assuming vanishing modular weights for Higgs fields.}
\end{longtable}
\end{center}

%% file: app_Majorana.tex
Following a similar procedure as in Appendix~\ref{app:classification_Dirac} with the discussion of section~\ref{sec:MajoranaYukawa}, we consider all Majorana mass textures. In this case, there are not as many cases as for Dirac masses. For example, if the Majorana term arises from a type-I seesaw mechanism, the number of inequivalent cases results from counting the combinations of possible $2D_3$ doublets for the neutrino fields $N_D^c$ and singlets for $N_3^c$. As there are two doublets and four singlets, there are only eight possible Majorana mass textures. It turns out that there are also only eight textures for Majorana masses generated via the Weinberg operator.
\enlargethispage{\baselineskip}

\begin{table}[b!]
\begin{center}
\resizebox{\textwidth}{!}{
\begin{tabular}{|c|c|c|c|}\hline\hline
	Model & \multicolumn{2}{c|}{Representation configuration} & \multirow{2}{*}{Majorana mass texture} \\ \cline{2-3}
	label & $N^c$ & $(L,(H_u)^2)$                                 &  \\
	\hline
	$\mathcal{N}_{1}$& $\qa\oplus\pa$ & $(\qa\oplus\pa,\pa)$, $(\qb\oplus\pc,\pb)$ & $\left(\begin{array}{ccc}
		\alpha_2 Y^{(k_{DD})}_\mathbf{1}-\alpha_1 Y^{(k_{DD})}_{\mathbf{2},1}~&~\alpha_1 Y^{(k_{DD})}_{\mathbf{2},2}~&~\beta Y^{(k_{D3})}_{\qa,1}\\
		\alpha_1 Y^{(k_{DD})}_{\mathbf{2},2}~&~\alpha_1 Y^{(k_{DD})}_{\mathbf{2},1}+\alpha_2 Y^{(k_{DD})}_\mathbf{1}~&~\beta Y^{(k_{D3})}_{\qa,2}\\
		\beta Y^{(k_{D3})}_{\qa,1}~&~\beta Y^{(k_{D3})}_{\qa,2}~&~\gamma Y^{(k_{33})}_{\pa}\\
	\end{array}\right)$ \\ \hline
	$\mathcal{N}_{2}$& $\qa\oplus\pb$ & $(\qa\oplus\pb,\pa)^+$, $(\qb\oplus\pd,\pb)^-$ & $\left(\begin{array}{ccc}
		\alpha_2 Y^{(k_{DD})}_\mathbf{1}-\alpha_1 Y^{(k_{DD})}_{\mathbf{2},1}~&~\alpha_1 Y^{(k_{DD})}_{\mathbf{2},2}~&~\mp\beta Y^{(k_{D3})}_{\qa,2}\\
		\alpha_1 Y^{(k_{DD})}_{\mathbf{2},2}~&~\alpha_1 Y^{(k_{DD})}_{\mathbf{2},1}+\alpha_2 Y^{(k_{DD})}_\mathbf{1}~&~\pm\beta Y^{(k_{D3})}_{\qa,1}\\
		\mp\beta Y^{(k_{D3})}_{\qa,2}~&~\pm\beta Y^{(k_{D3})}_{\qa,1}~&~\gamma Y^{(k_{33})}_{\pa}\\
	\end{array}\right)$ \\ \hline
	$\mathcal{N}_{3}$& $\qa\oplus\pc$ & $(\qa\oplus\pc,\pa)$, $(\qb\oplus\pa,\pb)$ & $\left(\begin{array}{ccc}
		\alpha_2 Y^{(k_{Y_{DD}})}_\mathbf{1}-\alpha_1 Y^{(k_{DD})}_{\mathbf{2},1}~&~\alpha_1 Y^{(k_{DD})}_{\mathbf{2},2}~&~\beta Y^{(k_{D3})}_{\qb,1}\\
		\alpha_1 Y^{(k_{DD})}_{\mathbf{2},2}~&~\alpha_1 Y^{(k_{DD})}_{\mathbf{2},1}+\alpha_2 Y^{(k_{DD})}_\mathbf{1}~&~\beta Y^{(k_{D3})}_{\qb,2}\\
		\beta Y^{(k_{D3})}_{\qb,1}~&~\beta Y^{(k_{D3})}_{\qb,2}~&~\gamma Y^{(k_{33})}_{\pb}\\
	\end{array}\right)$ \\ \hline
	$\mathcal{N}_{4}$& $\qa\oplus\pd$ & $(\qa\oplus\pd,\pa)$, $(\qb\oplus\pb,\pb)$ & $\left(\begin{array}{ccc}
		\alpha_2 Y^{(k_{DD})}_\mathbf{1}-\alpha_1 Y^{(k_{DD})}_{\mathbf{2},1}~&~\alpha_1 Y^{(k_{DD})}_{\mathbf{2},2}~&~-\beta Y^{(k_{D3})}_{\qb,2}\\
		\alpha_1 Y^{(k_{DD})}_{\mathbf{2},2}~&~\alpha_1 Y^{(k_{DD})}_{\mathbf{2},1}+\alpha_2 Y^{(k_{DD})}_\mathbf{1}~&~\beta Y^{(k_{D3})}_{\qb,1}\\
		-\beta Y^{(k_{D3})}_{\qb,2}~&~\beta Y^{(k_{D3})}_{\qb,1}~&~\gamma Y^{(k_{33})}_{\pb}\\
	\end{array}\right)$ \\ \hline
	$\mathcal{N}_{5}$& $\qb\oplus\pa$ & $(\qb\oplus\pa,\pa)^+$, $(\qa\oplus\pc,\pb)^-$ & $\left(\begin{array}{ccc}
		\alpha_2 Y^{(k_{DD})}_\mathbf{1'}\pm\alpha_1 Y^{(k_{DD})}_{\mathbf{2},2}~&~\pm\alpha_1 Y^{(k_{DD})}_{\mathbf{2},1}~&~\beta Y^{(k_{D3})}_{\qb,1}\\
		\pm\alpha_1 Y^{(k_{DD})}_{\mathbf{2},1}~&~\alpha_2 Y^{(k_{DD})}_\mathbf{1'}\mp\alpha_1 Y^{(k_{DD})}_{\mathbf{2},2}~&~\beta Y^{(k_{D3})}_{\qb,2}\\
		\beta Y^{(k_{D3})}_{\qb,1}~&~\beta Y^{(k_{D3})}_{\qb,2}~&~\gamma Y^{(k_{33})}_{\pa}\\
	\end{array}\right)$ \\ \hline
	$\mathcal{N}_{6}$& $\qb\oplus\pb$ & $(\qb\oplus\pb,\pa)^+$, $(\qa\oplus\pd,\pb)^-$ & $\left(\begin{array}{ccc}
		\alpha_2 Y^{(k_{DD})}_\mathbf{1'}\pm\alpha_1 Y^{(k_{DD})}_{\mathbf{2},2}~&~\pm\alpha_1 Y^{(k_{DD})}_{\mathbf{2},1}~&~-\beta Y^{(k_{D3})}_{\qb,2}\\
		\pm\alpha_1 Y^{(k_{DD})}_{\mathbf{2},1}~&~\alpha_2 Y^{(k_{DD})}_\mathbf{1'}\mp\alpha_1 Y^{(k_{DD})}_{\mathbf{2},2}~&~\beta Y^{(k_{D3})}_{\qb,1}\\
		-\beta Y^{(k_{D3})}_{\qb,2}~&~\beta Y^{(k_{D3})}_{\qb,1}~&~\gamma Y^{(k_{33})}_{\pa}\\
	\end{array}\right)$ \\ \hline
	$\mathcal{N}_{7}$& $\qb\oplus\pc$ & $(\qb\oplus\pc,\pa)^+$, $(\qa\oplus\pa,\pb)^-$ & $\left(\begin{array}{ccc}
		\alpha_2 Y^{(k_{DD})}_\mathbf{1'}\pm\alpha_1 Y^{(k_{DD})}_{\mathbf{2},2}~&~\pm\alpha_1 Y^{(k_{DD})}_{\mathbf{2},1}~&~\mp\beta Y^{(k_{D3})}_{\qa,2}\\
		\pm\alpha_1 Y^{(k_{DD})}_{\mathbf{2},1}~&~\alpha_2 Y^{(k_{DD})}_\mathbf{1'}\mp\alpha_1 Y^{(k_{DD})}_{\mathbf{2},2}~&~\pm\beta Y^{(k_{D3})}_{\qa,1}\\
		\mp\beta Y^{(k_{D3})}_{\qa,2}~&~\pm\beta Y^{(k_{D3})}_{\qa,1}~&~\gamma Y^{(k_{33})}_{\pb}\\
	\end{array}\right)$ \\ \hline
	$\mathcal{N}_{8}$& $\qb\oplus\pd$ & $(\qb\oplus\pd,\pa)^+$, $(\qa\oplus\pb,\pb)^-$ & $\left(\begin{array}{ccc}
		\alpha_2 Y^{(k_{Y_{DD}})}_\mathbf{1'}\pm\alpha_1 Y^{(k_{DD})}_{\mathbf{2},2}~&~\pm\alpha_1 Y^{(k_{DD})}_{\mathbf{2},1}~&~\beta Y^{(k_{D3})}_{\qa,1}\\
		\pm\alpha_1 Y^{(k_{DD})}_{\mathbf{2},1}~&~\alpha_2 Y^{(k_{DD})}_\mathbf{1'}\mp\alpha_1 Y^{(k_{DD})}_{\mathbf{2},2}~&~\beta Y^{(k_{D3})}_{\qa,2}\\
		\beta Y^{(k_{D3})}_{\qa,1}~&~\beta Y^{(k_{D3})}_{\qa,2}~&~\gamma Y^{(k_{33})}_{\pb}\\
	\end{array}\right)$ \\
	\hline\hline
\end{tabular}}
\caption{Majorana mass textures for Eq.~\eqref{eq:MajoranaSuperpotential} and Eq.~\eqref{eq:MajoranaSuperpotential_2} with $2D_3$ modular symmetry. We follow the same standards for the use of $\pm$ and $\mp$ as in Table~\ref{tab:DiracModels}. \label{tab:MajoranaModels}}
\end{center}
\end{table}

\enlargethispage{\baselineskip}
All possible Majorana mass textures in terms of the modular weights of the relevant VVMFs, as defined in section~\ref{sec:MajoranaYukawa}, are given in Table~\ref{tab:MajoranaModels}. The second column presents all different representation configurations for the right-handed neutrinos appearing in the type-I seesaw mechanism, Eq.~\eqref{eq:MajoranaSuperpotential}. The third column exhibits the configurations of $L$ and the product $H_u H_u$ in the Weinberg operator, Eq.~\eqref{eq:MajoranaSuperpotential_2}, considering all the different $2+1$-family structures for the field $L$ and every $2D_3$ singlet representation for the Higgs field $H_u$.